\documentclass{article}
\usepackage[tbtags]{amsmath}
\usepackage{amssymb}
\usepackage{amsthm}
\usepackage{array}
\usepackage{xy}

\usepackage[iso-8859-7]{inputenc}
\usepackage{graphicx}

\def\<{\left\langle}
\def\>{\right\rangle}

\title{\bfseries Gravitino dark matter production at finite temperature \normalfont}
\author{ \Large Iannis Dalianis \\
\\
\itshape Physics division, National Technical University of Athens, \\
\itshape 15780 Zografou Campus, Athens, Greece \normalfont\\
\\
\normalsize e-mail: dalianis@mail.ntua.gr }
\begin{document}

\date{}

\newcommand{\g}{\greektext} 
\newcommand{\e}{\latintext}
\maketitle

\abstract{The production and the final abundance of gravitino dark matter appear to depend crucially on the restoration of the global $U(1)$ $R$-symmetry of GMSB sectors in a threefold way. An $R$-symmetric phase effectively suppresses the production of goldstinos from scatterings with the supersymmetric Standard Model particles and it generically initiates the goldstino production from the thermalized messenger particles. In addition, the GMSB spurion gets displaced from the zero temperature minimum and under certain conditions it dominates the energy density of the universe producing late entropy. We show that it is possible to have high enough reheating temperatures that thermal leptogenesis and thermal vacuum selection can be realized without gravitino overproduction. The gravitino dark matter can be produced either thermally or non-thermally. In the former case the messenger scale has to be less than about $10^6$ GeV with the gravitino relatively heavy, $m_{3/2}\geq {\cal O}(10)$ GeV. In the later case, the gravitino is generically produced by the decay of the GMSB spurion field a process that always takes place for large messenger scales.  A connection of our results with current collider and observational data is performed.}
\\
\\
\tableofcontents

\section{Introduction} \normalsize

Supersymmetry is a very motivated candidate theory beyond the Standard Model of particle physics. 
Experimentally, we have not yet found any superpartners of the observed particles or profound supersymmetric deviations from the Standard Model. However, the discovery of a Higgs-like scalar at the Large Hadron Collider \cite{Aad:2012tfa, Chatrchyan:2012ufa} urges upon the hierarchy puzzle. 
The experimental data of the following years are expected to be of decisive importance for unrevealing the physics and symmetries of the TeV scale.

If the physics beyond the Standard Model is supersymmetric it is necessary to understand the effects of finite temperature in supersymmetric theories because they are expected to describe physics in the early universe. The supersymmetry introduces scalar degrees of freedom hence the vacuum structure of the theory becomes complex. The thermal effects can either drive phase transitions in the MSSM or in the gauge mediated supersymmetry breaking sector \cite{Craig:2006kx, Katz:2009gh, Dalianis:2010yk, Dalianis:2010pq, Hanken:2013gpa}.  Also, if the early universe experiences temperatures larger than the sparticle masses then the supersymmetric degrees of freedom will get excited  affecting the evolution of the universe and if longlived can be part of the dark matter. In supersymmetric theories the Lightest Supersymmetric Particle (LSP) can be stable. When the supersymmetry breaking is mediated by mainly Standard Model gauge interactions the LSP is generically predicted to be the gravitino and its present relic densisty has to be $\Omega_{3/2}h^2\lesssim \Omega_\text{DM} h^2 =0.118$ \cite{Ade:2013lta} where $h$ is the Hubble constant in units of $100\, \text{km} \, {\text{s}}^{-1}\, {\text{Mpc}}^{-1}$. 

The gauge mediation supersymmetry breaking (GMSB) scenario is well motivated from the particle phenomenology perspective, for a review see \cite{Giudice:1998bp}. 
In addition to the phenomenology, it can be a cosmollogically viable theory. It is free from Polonyi-like and gravitino problems due to the charged messenger sector that lies in an intermediate scale $M_\text{EW}\ll M_\text{mess}\ll M_\text{Pl}$.  Actually, in the GMSB context the gravitino is  a "problem" in the sense that the dark matter energy density observed at the universe  can be explained  in terms of thermally produced gravitinos only if a tuning is applied either at the reheating temperature (for not-equilibrated gravitinos) or the gravitino mass value (for gravitinos equilibrated with the MSSM fields). In the later case the gravitino mass has to be in the keV range and this value is disfavoured by the cosmological data \cite{Ade:2013lta, Boyarsky:2008xj}.

The gravitino is by definition a totally hidden sector sparticle. 
If the gravitino had only gravitational interactions the gravitino cosmology would be simple. However, this is true only for the helicity $\pm 3/2$  component. Once supersymmetry breaks the gravitino becomes massive via the superhiggs mechanism and a goldstino fermion accounts for the $\pm 1/2$ helicity component. For gravitino LSP the goldstino couplings with the MSSM  are enhanced compared to the $\pm 3/2$ helicity component. Moreover, the goldstino fermion can generically be coupled with other hidden sector fields. These imply that the gravitino dynamics may be subtle and the calculation of the final yield non-trivial. In most of the cases it is the reheating temperature or a temperature related to the hidden sector mass scales that controls the final yield albeit, it is very possible that non-thermal processes can be the dominant source.  Apparently, the gravitino LSP cosmology is radically different than the neutralino case which accounts for a visible sector sparticle that is part of the thermal equilibrium until the, comparetively low, freeze out temperature $T^f_{\chi^0}={\cal O}(10 \,\text{GeV})$.  

The calculation of the gravitino final yield is a compound task that involves knowledge of the MSSM properties, the behaviour of supersymmetry at finite temperature and the secluded/hidden sector dynamics.

\subsection{Overview}

The following superpotential is the basic paradigm of gauge mediation supersymmetry breaking:
\begin{equation} \label{min-3}
W=FX+\lambda X\phi \bar{\phi}\,,
\end{equation}
where $X$ is the Standard Model singlet spurion that breaks the supersymmetry, $\phi$, $\bar{\phi}$ the Standard Model charged messenger fields that mediate the supersymmetry breakdown to the observale sector and $\lambda$ is the messenger coupling; for review see ref. \cite{Giudice:1998bp}. Here we are working in the spurion limit, where the effects of supersymmetry breaking are encoded by the expectation values $\left\langle X \right\rangle=X_0+\theta^2 F_X$ and any microscopic dynamics can be neglected. Generally, we also assume the messenger sector is weakly coupled. This model has a particular global U(1) symmetry, the $R$-symmetry, under which $R[\phi]=R[\bar{\phi}]=0$ and $R[X]=2$. This symmetry is important in models of spontaneous $N=1$ supersymmery breaking \cite{Nelson:1993nf, Intriligator:2007py}. It is an accidental symmetry, since it has not been imposed on the theory but it is consequence of supersymmetry, gauge symmetry and field dimensionality. In order for the superpotential term $
\int d^4 x [W]_F$ to conserve $R$, the superpotential itself must have $R[W]=2$. The $R$-symmetry is broken spontaneously by the lowest component vev of $X$. However, the $X_0$ is not determined at the tree level. In the case of canonical K\"ahler and in the tree level approximation there are only supersymmetric vacua and the $X$  is a pseudomodulus, i.e. it is a flat direction with constant potential value $F^2$.  The Coleman-Weinberg potential from the interaction between messengers and the spurion  usually results in minimum at the origin that is not stable at the messenger direction.
 The spurion can be stabilized due to corrections to the K\"ahler potential for the spurion. A general expression is of the form
\begin{equation} \label{KSBR-3}
K=|X|^2+\epsilon_4 \frac{|X|^4}{\Lambda^2_*}+\epsilon_6\frac{|X|^6}{\Lambda^4_*}.
\end{equation}
where $\Lambda_*$ is a cut-off scale related  with the microscopic structure of the theory. For $\epsilon_4=1$ and $\epsilon_6<0$ the minimum lies at $X_0 \sim \Lambda_*/\sqrt{|\epsilon_6|}$ and the $R$-symmetry is spontaneously broken \cite{Shih:2007av, Lalak:2008bc, Dalianis:2010yk}.
In the case that there is a bare mass in the messenger sector apart from the vev of $X$ the $R$-symmetry may break explicitly. For example, in the superpotential (\ref{min-3}) we can add an $R$-violating messenger mass $\delta W=M \phi \bar{\phi}$. A dimensionful constant, $\delta W=c$, that cancels the potental energy of the phenomenologically acceptable vacuum is also an $R$-violating contribution. For $\epsilon_4=-1$ the minimum  in the $X$ direction lies at $X_0=\sqrt{3}\Lambda^2_*/(6M_\text{Pl})$ when $\delta W=c$ \cite{Kitano:2006wz} or, $X_0=-M/\lambda$ when $\delta W=M \phi \bar{\phi}$ \cite{Murayama:2006yf}.

Majorana gaugino masses are closely related to the vacuum structure of the theory, see e.g. ref. \cite{Intriligator:2007py, Komargodski:2009jf, ArkaniHamed:2004yi}. If supersymmetry breaking is not accompanied by $R$-breaking a split spectrum of gaugino and sfermion soft masses emerges. The $R$-symmetry forbids the appearance of Majorana gaugino masses, $m_\lambda$, $A$- and $\mu$-terms.
The cosmological implications of the thermal restoration of the $U(1)_R$ symmetry were mentioned in ref. \cite{Dalianis:2011ic}. Namely, an $R$-suppression of the $m_{\lambda}$ and $A$- soft terms correspondingly suppresses the helicity $\pm1/2$ gravitino component  production rate from MSSM thermal scatterings  which is controlled by the ratio ${m^2_{\lambda}}/{m^2_{3/2}}$. 
An $R$-symmetric phase may also affect the gravitino mass. In supergravity  a non zero value for the scalar component of the superpotential $\left\langle W \right\rangle \neq 0$ is neccessary for tunning the vacuum energy to a nearly zero value and this explicit cancellation of the cosmological constant breaks the $R$-symmetry, a breaking communicated to the visible sector by the gravitational interactions. It is this cancellation that gives a gravitino mass $m_{3/2}= e^{K/(2M^2_\text{Pl})}{|W|}/{M^2_\text{Pl}}$,
where $M_\text{Pl}=2.4 \times 10^{18}$ GeV is the reduced Planck mass. However, generally in an $R$-symmertic phase the ratio $m_{\lambda}/m_{3/2}$ is expected to be smaller than one. Indeed the scalar soft masses, $\tilde{m}$, contribute to the vacuum energy a fact that implies a lower bound for the gravitino mass while the Majorana gauginos mass can nearly vanish \cite{ArkaniHamed:2004yi}.

At high energies the goldstino, that appears once local supersymmetry breaks spontaneously,  corresponds to the helicity $\pm1/2$ gravitino component.
 The GMSB goldstino has a tree level coupling with the messenger fields which can generate goldstinos more efficiently than the MSSM fields. Indeed, for temperatures larger than the messenger mass scale, $M_\text{mess}$, the messenger fields number density have an unsuppressed thermal distribution and generate goldstinos via $2 \rightarrow 2$ scatterings with cross section scaling like $\lambda^2 /s$, where $s$ is the center of mass energy. 
  This cross section peaks when $T\sim M_\text{mess}$ and is generally large enough to bring the goldstinos into thermal equilibrium regardless the gravitino mass value. Moreover, the messengers at temperatures $T\sim M_\text{mess}$ will decay into goldstinos with a width $\Gamma_\lambda\propto \lambda^4 F^2_X/M^3_\text{mess}$. For sufficiently small messenger scales, $M_\text{mess}\lesssim 10^6$ GeV,  the  goldstino yield from messenger decays (this branching ratio may be negligible as we will show) can dominate over the one from scatterings. The conclusion is that once messengers acquire an equilibrium abundance the gravitinos get overproduced. Therefore, it seems that in the absence of any late entropy production the reheating temperature, $T_\text{rh}$, has to be less than the messenger mass and the natural, thermal selection of the supersymmetry breaking vacuum \cite{Dalianis:2010pq} as well as the thermal leptogenesis scenario \cite{Davidson:2008bu} cannot be realized.
   
A small value for the superpotential coupling $\lambda$ does decrease the final gravitino yield from messenger scatterings and decays. However, in order the dark matter constraint $\Omega_{3/2}h^2 \leq 0.11$ to be  fulfilled the messenger scale has to be small and the coupling ultra small, $\lambda \lesssim 10^{-10}$.  
Nevertheless, there is also another way to suppress the interaction strenght between the GMSB messenger fields and the goldstino fermion. The messengers couple with the $X$ superfield that may not be the only source of supersymmetry breaking. It is a possible scenario that the goldstino does  {\itshape not} reside in a single chiral superfield.  This is the case when the supersymmetry breaking scale is divided into the secluded GMSB and a hidden sector that mediates the breaking to the visible one only via gravity.  In the same fashion with the spurion of GMSB we can parametrize the supersymmetry breaking effects of this hidden sector by the vev of another SM singlet spurion superfield $Z$,  $\left\langle Z \right\rangle=Z_0+\theta^2 F_Z$. Therefore we have the relation
\begin{equation}
F^2=F^2_X+F^2_Z\,,
\end{equation}
where $F$ is the fundamental scale of supersymmetry breaking and $F_X$ is the scale of supersymmetry breaking felt by the messenger paticles. The $\lambda F_X$ is the mass splitting inside the messenger supermultiplets and its size dictates the critical temperatures of the secluded GMSB sector \cite{Dalianis:2010pq}. A direct consequence of the supersymmetry breaking scale partition is that the goldstino fermion becomes a linear combination of the the fermionic components $\psi_X$ and $\psi_Z$ of the superfields $X$ and $Z$ respectively.

The ratio $k\equiv F_X/F$ signifies whether the supersymmetry breaking is stronger in the gauge mediation or the gravity sector. Notwithstanding, we assume that the gauge mediation contribution to the SM sparticle massses is the leading one, hence, the gravitino mass 
\begin{equation}
m_{3/2}=\frac{F_X}{k\sqrt{3}M_\text{Pl}}
\end{equation}
is smaller than the radiatively generated soft masses. Otherwise the attractive features of the gauge mediation as the calculability and the absence of the flavour problem are lost. Actually this is a working assumption here for the gravitino is considered the LSP and dark matter candidate.

When $k \sim 1$ the spurion $X$ is the dominant source of supersymmetry breaking and the basic gauge mediation features remain intact. On the other hand, when $k \ll 1$ the extra dynamics of this hidden sector have to be taken into account\footnote{Except if $k$ is that small due to e.g. indirect communication of the supersymmetry breaking to the messenger superfields.}. Since this sector breaks supersymmetry as well we consider that it is also accompanied by a $U(1)_R$ symmetry that is preserved in the vacuum state. This way the $Z$-hidden sector vacuum becomes an enhanced symmetry point and thus attractive during the cosmological evolution.  Moreover, the gravity-mediated contributions to the sparticle spectum increase mainly the sfermion masses and not the $R$-violating gaugino masses. Another important remark is that in the $k \ll 1$ scenario the coupling of the gauge mediation messsenger fields with the goldstino field are suppressed because messengers couple only partially with the actual goldstino. Therefore, the goldstino production from the messenger fields is much less efficient and the gravitino overproduction problem from the thermalized messengers, as we will present, ameliorates or even gets solved for low messenger scale.

The estimation of the gravitino yield is controlled by the product of the messenger coupling $\lambda$ and the supersymmetry breaking parameter $k$. The observed dark matter density indicates that the $\lambda \cdot k$ takes small values.  However, the $\lambda\cdot k$  is related with the other supersymmetry breaking sector parameters $F$ and $M_\text{mess}$ by the expression 
\begin{equation} \label{barL}
\lambda k= \frac{M_\text{mess} }{F} \bar{\Lambda}
\end{equation}
and cannot become arbitrary small. The value of $\bar{\Lambda}$ determines the MSSM soft masses $\tilde{m}\sim (\alpha/4\pi)\bar{\Lambda}$ and it is tied to the electroweak scale -though constrained by the LHC data $\bar\Lambda \gtrsim 10^5$ GeV.  For fixed $\bar{\Lambda}$ a vanishing $\lambda\cdot k$  implies either a small messenger scale or a heavy gravitino. The parameter space that we consider here is the messenger scale to be $M_\text{mess} \geq 10^4$ GeV and the gravitino mass $m_{3/2} \leq 100$ GeV.

Nowadays the beyond the Standard Model particle physics has entered the LHC era. The discovery of a Higgs-like boson in the mass region around 125-126 GeV,  with a statistical significance at the level of 5-sigma, by the CMS and ATLAS teams have a significant impact on the supersymmetric theories. It is actually challenging to explain this Higgs boson mass in the context of minimal gauge mediation. In the minimal setup, MSSM plus GMSB sector described by (\ref{min-3}), the $m_h \approx 126$ GeV can be explained by large $A$-terms, above 1 TeV, generated through renormalization group evolution from high messenger scales \cite{Draper:2011aa}. Otherwise, the MSSM scalars are too heavy to be withn the reach at the LHC and the tension between the Higgs mass and MSSM naturalness deteriorates. Another possible scenario is the presence of marginal superpotential interactions between MSSM and messenger superfields. This way large $A$-terms can be generated directly at low messenger scales, see e.g. ref. \cite{Craig:2012xp, Evans:2013kxa}. 

It is an interesting result that both cases, with high and low messenger scale, that generate large $A$-terms can be cosmologically viable even if messengers are thermalized. The case that the messenegr scale, $\lambda X_0$, is high corresponds to relatively large values for the coupling $\lambda$ and goldstinos get overproduced. However, the spurion $X$ causes a late entropy production automatically and the dilution can have, under particular conditions, the correct magnitude \cite{Fukushima:2012ra}.  In the case that messenger scale is low the goldstino production from the messenger fields can be small enough that the goldstinos are genarated mainly by the MSSM fields without any dilution to interfere at low temperatures. We note that in the later case the gravitino production from the MSSM takes place at the characteristic temperature of the $R$-violating phase transition denoted by $T_{\not{R}}$.

Therefore high reheating temperatures are found not to be problematic in the GMSB scenarios. On the contrary, high reheating temperatures are welcome both for thermal leptogenesis and for the thermal selection  of the supersymmetry breaking vacuum \cite{Dalianis:2010pq}.

\subsection{Outline of the results}

The gravitino dark matter theory has been developed in several notable works. In the context of the MSSM the results of ref. \cite{Ellis:1984eq, Moroi:1993mb, Bolz:2000fu,Pradler:2006qh, Rychkov:2007uq} established the relation between the gravitino abundance and the reheating temperature. In the context of the GMSB models the messenger sector changes essentially the final gravitino yield from the thermal plasma. In the ref. \cite{Choi:1999xm} it was shown that for temperatures larger than the messenger mass the messenger sector contribution to the gravitino yield is much larger than that of the MSSM because the messenger  particles have stronger couplings to the goldstino and thus a stringent upper bound on the reheating temperature was found -apart from particular choices for the masses $m_{3/2}$ and $M_\text{mess}$. Moreover,  for viable cosmology the messenger fields cannot be stable  unless the lightest messenger particle has mass less than 5 TeV \cite{Dimopoulos:1996gy}. Hence the messengers, being unstable particles, can have a further impact on the cosmological evolution. If the lightest messenger decays slowly enough then it can be a source of a late entropy production diluting the gravitino and any pre-existing abundance \cite{Fujii:2002fv}. The decay width of the messenger particles depends on the messenger number violating interactions and hence the lightest messenger dilution magnitude can vary significantly \cite{Jedamzik:2005ir}. Another basic degree of freedom of the GMSB models is the spurion field. In the case that the spurion oscillations dominate the energy density of the universe then it will produce a late entropy among with non-thermally produced gravitinos. It is found in ref. \cite{Ibe:2006rc, Hamaguchi:2009hy, Fukushima:2012ra} that the yield of these non-thermal gravitinos is possible to be of the right order of magnitude.

{\itshape In this work} we re-investigate the physics of the gravitino production in the thermalized plasma of the early universe. We take into consideration that thermalized messengers induce a thermal mass on the spurion and for temperatures larger than the $T_{\not{R}}$, which is the characteristic temperature of the $R$-symmetry breaking, the spurion vev is shifted to the origin of the field space. Following the result of ref. \cite{Dalianis:2011ic} we assume that for $T>T_{\not{R}}$ the gravitino production from the MSSM plasma effectively vanishes. On the other hand the relativistic messengers have a thermal distribution and efficiently generate goldstinos.
We observe, actually, that for not too small superpotential coupling value, $\lambda$, the goldstinos generated by messenger scatterings acquire a thermal distribution. 

Motivated by recent phenomenological studies of GMSB models that suggest a low messenger scale with marginal superpotential couplings of messengers with the MSSM sector for interpreting the LHC data, e.g. ref. \cite{Craig:2012xp, Evans:2013kxa}, we focus on the low messenger scale scenario and we examine its cosmological implications. We find that for low messenger scale the gravitino can have an acceptable relic density, $\Omega_{3/2} \leq \Omega_\text{DM}$, only if the production of goldstinos from the thermalized messenger particles gets suppressed. This happens if the gravitino is relatively heavy $m_{3/2} > 1$ GeV and the superpotential coupling is small enough. The latter finding is also desirable for the thermal selection of the supersymmetry breaking vacuum \cite{Dalianis:2010pq}. The former, i.e. a heavy gravitino, indicates that there is an extra sector that breaks supersymmetry, but not the $U(1)_R$, and is not coupled with SM charged messengers fields. Thus the gravitino mass is $k^{-1}$ times enhanced despite the low value of the $M_\text{mess}$. Moreover, the marginal superpotential couplings of messengers with the MSSM render the lightest messenger shortlived and hence its decay does not dilute the thermal plasma. Therefore, if the messenger scale is low then the gravitinos get generated mainly by the MSSM plasma at the temperature $T_{\not{R}}$. The cosmological constraint for the gravitinos, $\Omega_{3/2} \leq \Omega_\text{DM}$, can be satisfied in a part of the parameter space where $10^4\, \text{GeV}< M_\text{mess} \lesssim 10^6$ GeV, $m_{3/2}> 1\, \text{GeV} $ and reheating temperatures $T_{\not{R}}<T_\text{rh}<T_{\pm3/2}$. The $T_{\pm3/2}$ stands for the temperature that the helicity $\pm3/2$ gravitino component, generated via gravitational interactions, saturates the $\Omega_\text{DM}$ bound. We note that values for the messenger mass $M_\text{mess} < 10^5$ GeV can be reached only for $\bar{\Lambda}<10^5$ GeV which is a rather constrained possibility according to the LHC data.

There is a variance with the results of ref. \cite{Choi:1999xm} which concerns the goldstino yield from the messenger decays.
Here, we find that the branching ratio of the messengers to goldstinos is generically too small to have a leading contribution to $\Omega_{3/2}$ because we consider superpotential couplings of the messengers with MSSM superfields.

Above, we have implicitly assumed that the spurion $X$ will reach the zero temperature minimum without large oscillations. However, this cannot be always the case. Actually, we find that this generically happens at low messenger scales. For larger values of the messenger scale, $M_\text{mess}\gtrsim 10^{10}$ GeV, the spurion oscillations always dominate the energy density of the universe and dilution takes place. We mention that in the scenarios where the lightest messenger decays slowly enough to dilute the plasma, as described in ref. \cite{Fujii:2002fv}, the spurion afterwards reheats the universe and therefore, the messenger dilution effects can be generally neglected.

There are also phenomenological GMSB models that suggest a large messenger scale in order to explain the 126 GeV Higgs mass \cite{Draper:2011aa}. These models predict large mass for the gluinos $m_{\tilde{g}} \gtrsim 3$ TeV. However, as we explained, large masses for the gluinos do not deteriorate the gravitino cosmology due to the fact that in the large messenger scale case the gravitinos are produced mainly non-thermally by the spurion decay. In this work we point out the part of the parameter space where the non-thermally production of gravitinos takes place without carrying out any detailed calculation of the non-thermal gravitinos abundance. This can be found in the ref. \cite{Ibe:2006rc, Hamaguchi:2009hy, Fukushima:2012ra}. 

The article is organized as follows. In section 2 we discuss the behaviour of the supersymmetry at finite temperature mentioning issues as the vacuum selection and the values of the soft parameters in the reheated early universe. We also mention the important effects of an $R$-symmetric thermal phase reserving a more detailed discussion with explicit GMSB models for the appendix A. The section 3 contains the interaction terms of the gravitino and the goldstino superfields with the MSSM and the messenger sector. There we give the relevant cross sections and decay rates for the calculation of the final gravitino relic density. In section 4 the total gravitino yield from MSSM and messenger sector is computed taking into account possible late time production. This section contains the main results of this work. Complementary calculations and some technical details can be found in the appendix B. In section 5 we briefly comment on supersymmetry breaking theories where the $R$-symmetry restoration can be possibly realized without the thermalization of messenger fields; such theories would actually give a different gravitino cosmology. In section 6 we combine our results on the gravitino relic density with GMSB phenomenology of the LHC era. We also discuss complementary cosmological constraints as the large scale structure in the Universe, the Big Bang Nucleosynthesis (BBN) and the leptogenesis scenario. In the last section we conclude. 

\section{Supersymmetry breaking at finite temperature}

At finite temperature the fields that interact with the thermal plasma are no longer in their vacuum state. The occupation numbers are given by the Bose-Einstein formula. The temperature dependent one-loop effective potential is of the form   $V_1^T \sim T^4 \int dx\,x^2 \ln \left\{1\pm \exp \left(-\sqrt{A_i} \right) \right\}$
where $A_i=x^2+M^2_i/T^2$ and $M^2_i$ is an eigenvalue of the mass squared matrix \cite{Kolb:1990vq, Mukhanov:2005sc, Quiros:1999jp, Dalianis:2010yk}.

Knowledge of the behaviour of the supersymmetric field theory at finite temperature is basic for reliable cosmological calculations. The gravitino cosmology is a representative case of a such calculation.  Its relic abundnace depends on the precise values that the scale of supersymmetry breaking, the soft masses, the messenger scale and hidden  sector couplings have in a thermal phase. A particular example is the thermal restoration of the $R$-symmetry that appears to be a generic phenomenon in GMSB models.  
In the ref. \cite{Dalianis:2011ic} it was shown that the gravitino production from the MSSM sector is fairly insensitive to the reheating temperature once $R$-symmetry restoration takes place. In this work, we also include the contribution from the thermalized messenger sector to the final gravitino yield.
Another important issue tightly connected with the gauge mediation models  is the metastability of the supersymmetry breaking vacua. The vacuum structure of the supersymmetry breaking potential at finite temperature has been studied in ref. \cite{Dalianis:2010yk} where the danger of transition to a supersymmetric vacuum was underlined. 

High reheating processes as the leptogenesis or thermal gravitino production with mass in the MeV-GeV range and $\Omega_{3/2}\sim 0.2$ cannot be secure if the phenomenologically acceptable minimum is thermally unstable. 
\begin{figure} \label{X}
\centering
\includegraphics [scale=0.8, angle=0]{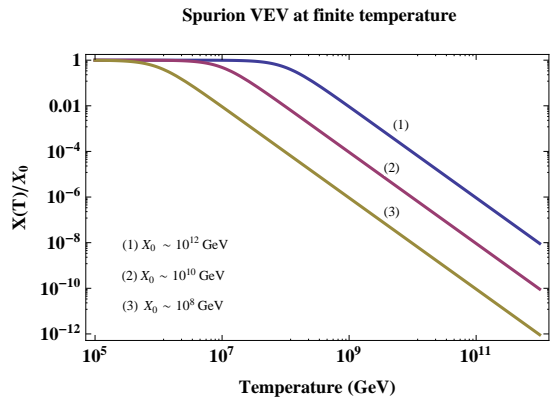}  
\caption{The figure shows the thermal evolution of the spurion vev, $X(T)$, over the zero temperature vev $X_0$ for different values of the cut-off scale $\Lambda_*=10^{-2}, \, 10^{-3}, \, 10^{-4} \, M_\text{Pl}$ (from top to bottom) and for superpotential coupling value $\lambda=10^{-5}$. The model considered is the (\ref{min-3}) plus $\delta W=c$ and $\delta K= |X|^4/\Lambda^2_*$.
It is apparent that as the temperature increases the spurion thermal average value approaches the origin of the field space.}
\end{figure}

\subsection{Thermally safe models of supersymmetry breaking}
By construction the gauge mediation scenario addresses the Polonyi cosmological problem thanks to the Yukawa couplings of the spurion with the messengers; however, it is these particular couplings that render the supersymmetry breaking vacuum \itshape metastable \normalfont. 
For the messenger superpotential, $W =\lambda X \phi \bar{\phi}$, the following vacuum selection conditions have been found \cite{Dalianis:2010pq}:
\begin{enumerate}
	\item Canonical K\"ahler for the spurion $X$ (regardless the value of $\lambda$): \itshape thermally "discarded". \normalfont 
	\item $X$-dependent corrections in the K\"ahler  and $\lambda \ll 1$: \itshape thermally favourable. \normalfont
\end{enumerate}

Thermally discarded means that, once the messengers get thermalized, the system of the fields evolves towards the supersymmetric minimum. In the thermally favourable case the supersymmetry breaking minimum is always more attractive. For the sake of comprehension we can demonstrate this interesting behaviour by assuming, as a gedunken experiment, that we heat a part of the universe up. If the heating temperature is less than the $M_\text{mess}$ then both theories, 1 and 2, behave in a similar manner. On the contrary, when the temperature exceeds the messengers mass the first class of theories will exhibit a thermal phase transition towards a supersymmetric state while the second class will exhibit a kind of "elastic" behaviour where the initial zero temperature state is restored after the cooling of the system\footnote{If we knew that somewhere in the observable universe collisions with (the ultra high) center of mass energy $\sqrt{s} \geq M_\text{mess}$ had taken place and an instantaneous local thermal equilibrium could have been realized then we could discard the first class of theories for describing the physics beyond the Standard Model. Otherwise, a bubble with the supersymmetric global minimum could have been generated.}. It is thus illustrative to characterize the first class of theories fire-sensitive and the second class fireproof.
\begin{figure} \label{VX}
\centering
\begin{tabular}{cc}
{(a)} \includegraphics [scale=0.8, angle=0]{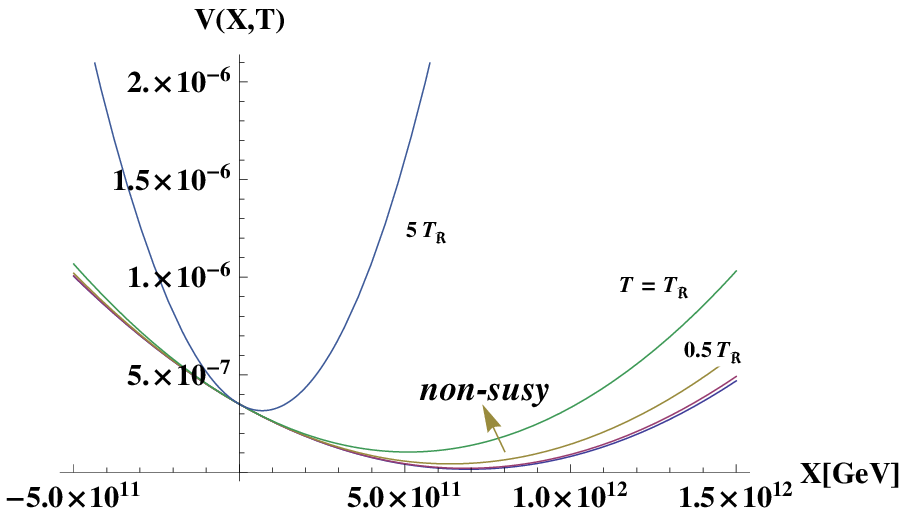}  &
{(b)} \includegraphics [scale=0.8, angle=0] {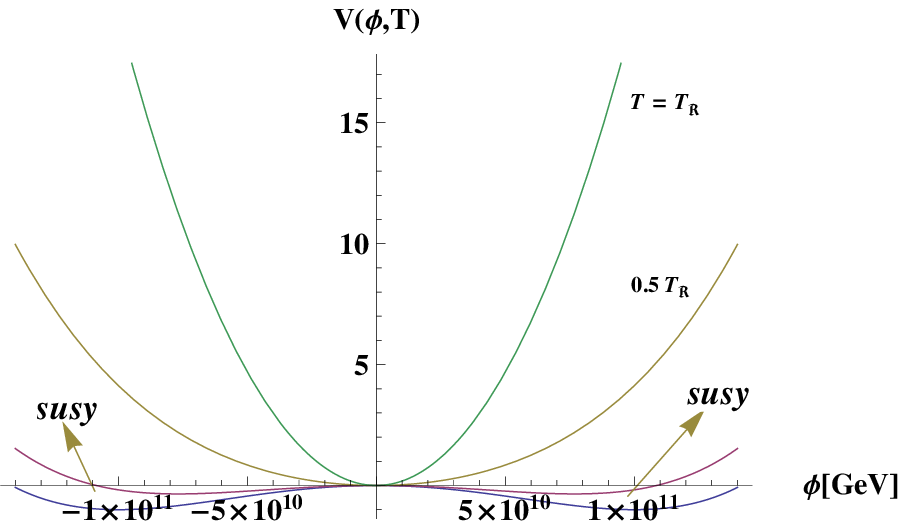}  \\
\end{tabular}
\caption{ The effective potential for the spurion field (left panel) and messenger fields (right panel) at the temperatures $5T_{\not{R}}, T_{\not{R}}, 0.5 T_{\not{R}}, 0.2T_{\not{R}}$ and $T=0$ from top to bottom. In the left panel, the breaking of the $R$-symmetry is manifest for temperatures $T < T_{\not{R}}$ : $X({T<T_{\not{R}}}) \approx X_0$. At the same temperature, $T_{\not{R}}$, in the right panel the second order phase transition has not yet taken place. We used the values $\Lambda_* =10^{-3}M_\text{Pl}$ for the cut-off, $\lambda=10^{-5}$ for the coupling and $F=10^{17} \text{GeV}^2$ for the susy breaking scale squared. }
\end{figure}

\subsection{Finite temperature effects on the supersymmetry breaking sectors}
Supersymmetry behaves differently from other symmetries. 
Exact supersymmetry at zero temperature breaks at $T\neq 0$ \cite{Das:1978rx}; this is a consequence of different 
statistics for bosons and fermions \cite{Girardello:1980vv}.  If the auxiliary fields acquire a nonvanishing thermal average a zero-momentum fermionic (collective) excitation can be created with zero energy. This massless excitation couples to the supercurrent in analogy with the Goldstone theorem at $T=0$
and vanishes as $T \rightarrow 0$ \cite{Boyanovsky:1983tu, Aoyama:1984bk}. 

In reference \cite{Leigh:1995jw} it was shown that the thermal Goldstone boson does not mix with the zero temperature goldstino and hence, cannot play any role in the gravitino production. Furthermore, the authors of ref. \cite{Ellis:1995mr} examined the leading thermal correction to the goldstino-matter coupling,
and found that these are negligible for $T \ll \sqrt{F}$ hence, the zero temperature estimates are qualitative correct. 
Here, in what concerns the soft masses at fininte temperature, we will consider the effects of the $R$-symmetry restoration which takes place at $T_{\not{R}}=4kF/(\lambda \Lambda_*\sqrt{N})$, a temperature not far away from the $\sqrt{F}$ scale.
\\
\\
{\itshape Goldstino - MSSM decoupling at high temperatures}\\
As we will expose in detail in the next section, the goldstino field, which at high energies accounts for the helicity $\pm 1/2$ gravitino component, couples stronger with MSSM gauge superfields rather than the chiral ones. In particular, the goldstino coupling to gauginos is proportional to $m_{\lambda}T/F$ and dominates over the coupling to sfermions which is proportional to $m^2_{\tilde{f}}/F$. The thermal effects can induce extra and possibly dramatic modification to the goldstino coupling. First of all it is the suppression of the gaugino masses due to $R$-symmetry restoration. Indeed, the finite temperature effective potential for the scalar component of the GMSB spurion superfield  has a vacuum  structure (illustrated in figures 1 and 2) that differs significantly from the zero temperature potential. It was found in ref. \cite{Dalianis:2011ic} that for $T>T_{\not{R}}$ the minimum of the effective potential preserves the $R$-symmetry a fact that implies vanishing gaugino masses according to the relation 
\begin{equation}
\frac{m_\lambda (T)}{m_\lambda} \sim \frac{X(T)}{X_0} =\frac{1}{1+\left( \frac{T}{T_{\not{R}}}\right)^2}\equiv b_R(T)\,.
\end{equation}
Due to the zero temperature relation $m_\lambda =(\alpha/4\pi)F_X/X_0$ one may conclude that the gaugino masses, and generally the soft masses, increase like $1/X(T)$ as the $X(T) \rightarrow 0$ . However, we expect that at finite temperature the soft masses do not to blow up as the $X$ field approaches the origin thanks to the collective phenomena of the thermal bath. In particular, the soft masses are generated via the messenger fields that run in the loops. At the zero temperature the messengers mass is $\lambda X_0$ whereas for $X(T)\ll X_0$ the messenger fields are relativistic and a plasmon mass $\sim gT$ for messengers has to be considered. From another perspective, we expect that the interaction range of the relativistic messengers is not infinite but it is screened in a Debye fashion. Apparently, the exact value of the soft masses is a non-trivial issue and we will not enter into more details here for this is out of this work's scope. From these considerations we can infer that on the one hand, sfermion masses remain finite at high temperature due to the non-zero plasmon mass for messengers, - taking also into account a possible thermal corrected $F$-term for $T \gtrsim \sqrt{F}$ - and, on the other hand, the gaugino masses are suppressed due to the $R$-symmetry restoration. 

Furthermore, we consider the gravitino mass to remain roughly unsuppressed in an $R$-symmetric phase because the superpotential value at finite temperature, $W(T)$, does not vanish. These conclusions are an  heuristic, though essential, rule for a consistent calculation of the final gravitino yield.

We finally comment on the variance found in the literature concerning the goldstino-MSSM interaction for temperatures larger than the messenger mass. The authors of ref. \cite{Choi:1999xm, Fukushima:2012ra} argued that for energies $T \gtrsim  M_\text{mess}$ 
there are no direct local couplings of the goldstino, $\psi_G$, to the MSSM particles at Lagrangian level and the $\psi_G$ decouples. On the contrary, the authors of ref. \cite{Jedamzik:2005ir} claimed that in local supersymmetry there are tree level particle-sparticle-goldstino interaction terms hence the goldstino does not decouple from the MSSM for temperatures $M_\text{mess} \lesssim T$ and, thus, the estimation of the gravitino abundance is sensitive to the reheating temperature. Here we follow the approach of the authors of ref. \cite{Jedamzik:2005ir}. The effective Lagrangian for the gravitino is obtained from the $N=1$ supergravity theory regardless the mechanism employed in the messenger sector to communicate the supersymmetry breaking to the visible sector \cite{Cremmer:1982en}.
 In our work, the new feature is that we take into account the $R$-symmetry dynamics and, as we argued, the MSSM-goldstino couplings get suppressed above the $T_{\not{R}}$ temperature.

Finally, for the sake of consistency, we mention that in this work we assume that all the gauge and superpotential couplings remain perturbative up to the reheating temperature, $T_\text{rh}$.

\section{Gravitino interactions}
The gravitinos are spin 3/2 particles and described by the Lagrangian for vector spinor fields. Its interactions can be read by the general N=1 supergravity Lagrangian (see e.g. \cite{Drees:2004jm, Moroi:1995fs}) which contains interaction terms between the gravitino field $\psi_\mu$ and the Noether current of local supersymmetry.

The gravitino acquires mass by the super-Higgs mechanism. For energies much larger than the gravitino mass and for $m_{3/2}$ not larger than the effective mass splitting in the supermultiplets the helicity $\pm1/2$ components of the gravitino can be treated as the true goldstino modes. This is implied by the equivalence theorem and the $\pm1/2$ gravitino component can be written for $\sqrt{s} \gg m_{3/2}$
\begin{equation} \label{gold}
\psi_\mu \sim i\sqrt{\frac23} \frac{1}{m_{3/2}}\partial_\mu\psi_G
\end{equation}
where $\psi_G$ denotes the goldstino. This is the case we study: TeV scale mass splitting in the observable sector supermultiplets with the gravitino the lightest sparticle with $m_{3/2} <100\, \text{GeV}$.

The goldstino field is the fermionic component of a chiral superfield that breaks supersymmetry. Here this role is played by the spurion $X$ which obtains a non-vanishing vacuum expectation value $\left\langle  X \right\rangle=X_0+\theta^2 F_X$. Still, the $X$ field may not be the only field that breaks supersymmetry. Other hidden sector fields, that we collectively denote them with $Z$, may have $F_Z\neq 0$ and the total $F$-term breaking value in such a case is  $F=\sqrt{F^2_X+F^2_Z}$. Hence, the gravitino mass reads $m_{3/2}={F_X}/{k\sqrt{3}M_\text{Pl}}$ where $k\equiv F_X/F$.
Correspondingly, the goldstino fermion is a linear combination of the the fermionic components $\psi_X$ and $\psi_Z$ of the superfields $X$ and $Z$ respectively:
\begin{equation}
\psi_G=\frac{F_X}{F}\psi_X+\frac{F_Z}{F}\psi_Z.
\end{equation}
We are going to consider both $k\sim 1$ and $k \ll 1$ scenarios.

\subsection{MSSM sector}
In the four-component notation and in the flat space time the gravitino interaction Lagrangian with chiral $(\tilde{f}^i, f^i)$ and gauge supermultiplets  $(A^a_\mu,\lambda^a)$ reads 
\begin{equation}\label{Gold-9}
{\cal L}_{eff,\,\text{goldstino}} =- \frac{i m_\lambda}{m_{3/2}} \frac{1}{8\sqrt{6}M_\text{Pl}} [\gamma^\mu, \gamma^\nu] \bar{\psi}_G \lambda^a F^a_{\mu\nu} + \frac{m^2_{\tilde{f}}-m^2_f}{m_{3/2}}\frac{1}{\sqrt{3}M_\text{Pl}} \bar{\psi}_G f_R \tilde{f}^* +\text{h.c.} \,.
\end{equation}
It is manifest from the above Lagrangian that the goldstino decouples from the theory in the supersymmetric limit. For energies $\sqrt{s}>m_\lambda, m_{\tilde{f}}$ single gravitinos can be created or annihilated dominantly through processes that include gauge superfields components $x,y,z$ with cross section
\begin{equation} \label{crossSM}
\sigma_{tot} =\frac{1}{2}\sum_{x,y,z} \eta_x\eta_y \sigma(x+y\rightarrow z+\psi_G)\simeq \frac{2.4g^2_1m_{\tilde{b}}^2+9.2g^2_2m_{\tilde{w}}^2+26g^2_3 m^2_{\tilde{g}}}{24\pi m^2_{3/2}M^2_\text{Pl}}
\end{equation}
where $m_\lambda$ (${\lambda=\tilde{g}, \tilde{w}, \tilde{b}})$ the supersymmetric Standard Model gaugino masses and $\eta_{x,y}$  equals 1 (3/4) for an incident boson (fermion). We notice that for $m_\lambda \rightarrow 0$ and $m_{\tilde{f}}\neq 0$ the cross section vanishes i.e. the goldstino appears to decouple from the MSSM without having supersymmetry restoration. This is the case of $R$-symmetry restoration.

Since the gravitino is the Lightest Supersymmetric Particle (LSP) the heavier sparticles will decay to a Standard Model field plus a gravitino. The decay width into gravitinos is nearly the same for both gauginos and sfermions
\begin{equation} \label{ssmd}
\Gamma^\text{MSSM}_\text{dec}(\tilde{i} \rightarrow i\,\psi_G  )\simeq\frac{1}{48\pi}\frac{\tilde{m}_i^5}{m^2_{3/2}M^2_\text{Pl}}
\end{equation}
where $\tilde{i}=\lambda, \tilde{f}$.

Apart from the interactions of the goldstino field with the Supersymmetric Standard Model (\ref{Gold-9}) it is crucial to include all the other interactions that may alter the final gravitino yield. Thereby, investigating the entire goldstino couplings is of central importance. 

\subsection{Messenger sector}
The supersymmetry breaking is communicated to the visible low energy sector via the messengers superfields. When these fields are charged under the Standard Model we have the ordinary gauge mediation scenario; when the breaking is communicated via gravity the $M_\text{Pl}$ suppressed interactions induce soft masses of order $m_{3/2}$.  We require that the GMSB messengers come in complete, vector-like $SU(5)$ representations and the SM gauge couplings remain perturbative up to the GUT scale. Messengers acquire supersymmetry violating masses from the non vanishing vacuum expectation value of the auxiliary component of the $X$ field via the superpotential coupling $\lambda X\phi\bar{\phi}$. Due to this coupling the Lagrangian contains the interaction term between the fermion $\psi_X$ and the messenger fields, $\delta {\cal L}_{X\phi\bar{\phi}} = -\lambda (\psi_X \chi_{\bar{\phi}}\phi + \psi_X \chi_{{\phi}}\bar{\phi} + \text{h.c.})$,
where $\chi_{\phi}$, $\chi_{\bar{\phi}}$ the fermionic components of the $\phi$, $\bar{\phi}$ messengers. The fermionic component of $X$ can be written in   terms of the goldstino fermion as $\psi_X=(F_X/F)\psi_G+...$ and the interaction 
is cast in
\begin{equation} \label{mess2}
\delta {\cal L}_{G\phi\bar{\phi}} = -\lambda k (\psi_G \chi_{\bar{\phi}}\phi + \psi_G \chi_{{\phi}}\bar{\phi} + \text{h.c.}).
\end{equation}
The scattering cross section for single goldstino production of messengers $\phi, \phi^*$ and plasmons for energies $\sqrt{s}\gg M_\text{mess}$ is given by \cite{Choi:1999xm}
\begin{equation} \label{crossME}
\sum_{A,A',B} \left[\sigma(A_\text{mess}+A'_\text{mess} \rightarrow B_\text{MSSM} + \psi_G)+ \sigma(A_\text{mess} + B_\text{MSSM} \rightarrow A'_\text{mess} + \psi_G) \right]\simeq\, 6\,\xi \lambda^2 k^2\frac{1}{s}
\end{equation}
where $A_\text{mess}=\phi, \chi_\phi$ are messenger field components, $B_\text{MSSM}= \lambda^a, V^a_\mu$ are gauge superfield components and $\xi=(g_1^2+3g_2^2+8g_3^2)/4\pi$. The above cross section has an inverse dependence on the center of mass energy squared  and  it maximizes for $s\simeq M_\text{mess}$ which is the threshold between relativistic  and non-relativistic, i.e. Boltzmann suppressed,  messenger particles (for further analysis see appendix B). The messenger cross section (\ref{crossME}) has a different behaviour than the cross section of the goldstino production  from MSSM fields, eq. (\ref{crossSM}), which is independent of the center of mass energy $\sqrt{s}$ and yields a goldstino abundance with a linear dependence on the temperature.

The lightest messenger particles may decay late or even be stable and contribute to the dark matter abundance. Their final abundance is determined by the thermally averaged annihilation cross section which for temperatures $T\ll M_\text{mess}$ 
takes the approximate form $\left\langle \sigma_\text{ann} v \right\rangle =a+6b/x_\text{mess} +{\cal O}(x^{-2}_\text{mess})$ where $x_\text{mess} \equiv M_\text{mess}/T$.
 Precisely, the thermally averaged annihilation cross sections to MSSM particles and goldstinos are parametrized like
\begin{equation} \label{sigmas}
\left\langle \sigma (\phi\phi^* \rightarrow \text{MSSM} )v\right\rangle=\frac{1}{M^2_\text{mess}} \left(A- \frac{B}{x_\text{mess}} \right) \,,\,\,\,\,\quad\quad
\left\langle \sigma (\phi\phi^* \rightarrow \psi_G\psi_G)v\right\rangle=\frac{k^4 \lambda^4}{M^2_\text{mess}} \left(A'-\frac{B'}{x_\text{mess}} \right)
\end{equation}
with $A \simeq B \simeq 3\times 10^{-3}$ 
and $A'=1/32\pi,\, B'=15/128\pi$ \cite{Dimopoulos:1996gy, Jedamzik:2005ir}.

The decay width of messenger particles to goldstinos is dominated by the superpotential coupling (\ref{mess2}) and reads
\begin{equation} \label{Gmess}
\Gamma_\lambda (\chi_\phi \rightarrow  \phi \, \psi_G) = \frac{1}{4\pi} \lambda^4 k^4 \frac{F^2}{M^3_\text{mess}}\, ,
\end{equation}
Further couplings of the messengers with the other sectors and in particular the observable sector are possible and have been proposed in the literatute with either cosmological or phenomenological orientation, see e.g. \cite{Jedamzik:2005ir} or \cite {Evans:2013kxa} respectively. Superpotential interactions are cosmologically expected in order to prevent the lightest messenger from being stable.  
Furtermore, a Higgs at 126 GeV it is hard to be explained in the minimal versions of gauge mediation plus MSSM; introducing marginal superpotential interactions between the messengers and the Higgses can ameliorate or even solve the problem. The cosmological implications of these extra messenger interactions, regarding the calculation of the gravitino abundance, will be discussed in the 4.2.2 subsection.

\subsection{Further hidden sector structure}

In this work we also investigate the scenario that the GMSB sector has a subleading contibution to the fundamental supersymmetry breaking scale, $F$, which is mainly sourced by the $Z$-hidden sector i.e. $F\simeq F_Z$. Respectively, the goldstino fermion is mainly composed by a $Z$-fermion. Generally, the gravitational couplings between the gravitino and the other hidden fields $Z_i$ can be derived from the gravitino mass term $ {\cal L} =-e^{G/2} \left(\psi_\mu \sigma^{\mu\nu} \psi_\nu +\text{h.c.} \right) $ where $G=K+\text{ln}(W^*W)$ and we have set $M_\text{Pl}=1$, see e.g. \cite{Dine:2006ii}.  After the Taylor series expansion of the exponential we get ${\cal L} = -\frac{1}{2} \left\langle e^{G/2} G_i \right\rangle Z_i \psi_\mu \sigma^{\mu\nu}\psi_\nu$. 
If the number density of these hidden sector particles in the early universe is large enough and the branching ratio to gravitinos is significant then the $Z$-hidden sector can be cosmologically problematic.  The study of non-thermal production of gravitnos from other than the GMSB secluded sector is very model dependent and out of the scope of this work. Here we assume that the $Z$-sector is naturally equiped with an exact $U(1)_R$ symmetry \cite{Nelson:1993nf} which is important for justifying that the $Z_i$ scalars are not displaced from their zero temperature minimum in the post-inflationary era.  Therefore the SM gauge singlet scalar, $Z$, which has no Yukawa couplings to messengers may not cause a Polonyi-like cosmological problem and not overproduce gravitinos via its decay.

\section{Gravitino relics from the thermalized plasma}

The gravitino number density $n_{3/2}$ in the thermalized early universe evolves according to the Boltzmann equation \cite{Kolb:1990vq}
\begin{equation}
\frac{dn_{3/2}}{dt}+ 3H n_{3/2} = \left(\gamma_\text{sc}+\gamma_\text{dec}\right) \left( 1-\frac{n_{3/2}}{n^\text{eq}_{3/2}}   \right).
\end{equation} 
The $\gamma_\text{sc}=\gamma^\text{MSSM}_\text{sc}+\gamma^\text{mess}_\text{sc}$ is the gravitino production rate in scatterings with thermalized Standard Model sparticles and messengers and reads
\begin{equation}
\gamma_\text{sc}= 0.14 \frac{T^6}{M^2_\text{Pl}} \left(1+\frac{b^2_R(T) m^2_{\tilde{g}}(\mu)}{12 m^2_{3/2}} \right)+  \xi \lambda^2 k^2 \frac{T^6}{s} f(s)\,,
\end{equation}
for six effective quark-squark flavour multiplets and $m_{\tilde{g}}(\mu)=m_{\tilde{g}}(T)g^2(\mu)/g^2(T)$ with $\mu$ about the electroweak scale. The $f(s)$ is a dimensionless function of the center of mass energy squared that can be found in the appendix.
The $\gamma_\text{dec}$ accounts for the gravitino production rate from decays of particles in thermal equilibrium 
\begin{equation}
\gamma_\text{dec}=\sum^{N}_{i=1} n^\text{eq}_i \frac{m_i}{\left\langle E\right\rangle} \Gamma_\text{dec}= \sum^{N_\text{MSSM}}_{j=1} n^\text{eq}_j \frac{m_j}{\left\langle E\right\rangle} \Gamma^\text{MSSM}_\text{dec} +\sum^{N_\text{mess}}_{l=1} n^\text{eq}_l \frac{m_l}{\left\langle E\right\rangle} \Gamma^\text{mess}_\text{dec}
\end{equation}
where $m_i/\left\langle E\right\rangle$ is the thermal average of the time dilation factor, $m_j \simeq \tilde{m}$ the soft masses and $m_l \simeq M_\text{mess}$. We mention that the messenger particles share roughly the same mass for small messenger mass splittnigs, i.e. $M^2_\text{mess} \gg \lambda F$. The decay rate of the Standard Model sparticles into gravitinos is given by the eq. (\ref{ssmd}) and that of the messengers by the (\ref{Gmess}).

\subsection{Thermal production of gravitinos}

\subsubsection{MSSM sector}
In a supersymmetric thermal bath 
 the gravitino possesses only the helicity $\pm3/2$ components and interacts with $M_\text{Pl}$ suppressed interactions. A thermal distribution for gravitationally interacting gravitinos can be achieved only for unrealistic temperatures  $T>T^f_{3/2}\sim {\cal O} (M_\text{Pl})$. In a softly broken supersymmetric thermal bath the longitudinal mode of the gravitino interacts with the observable sector with strength enhanced by $m_{\lambda}/m_{3/2}$.  This could result in the thermalization of the gravitinos by the $a+b \leftrightarrow c+ \psi$ scatterings in the MSSM plasma giving a freeze out temperature
\begin{equation}
T^{\text{eq, MSSM}}_{3/2} \sim 2\times 10^{10} \text{GeV} \left( \frac{m_{3/2}}{10 \text{MeV}} \right)^2 \left(\frac{1 \text{TeV}}{m_{\tilde{g}}} \right)^2.
\end{equation}
However, this is true only if $T^{\text{eq, MSSM}}_{3/2}<T_{\not{R}}$, which is usually not the case, unless the gravitinos are very light. For gravitino mass above the MeV range the $T_{\not{R}}$ is smaller than the $T^{\text{eq, MSSM}}_{3/2}$ and the gravitinos do not equilibrate with the MSSM plasma.
The gravitino abundance in units of the critical density from scatterings with the MSSM plasmons for reheating temperatures $T_{rh}>T_{\not{R}}$ is given by the expression \cite{Dalianis:2011ic}
\begin{equation} \label{OgA}
\Omega^\text{MSSM (th)}_{3/2}h^2 \simeq 0.1 \left(\frac{\theta_{rh} T_{\not{R}}}{10^8\,\text{GeV}} \right) \left(\frac{\text{GeV}}{m_{3/2}} \right) \left(\frac{m_{\tilde{g}}(\mu)}{1\, \text{TeV}} \right)^2\,
\end{equation}
where  $\theta_{rh}\sim 3/2$ for $T_{rh}>10T_{\not{R}}$, $\mu \sim \, 100\, \text{GeV}$ for the running of gluino mass $m_{\tilde{g}}$ from the electroweak scale to the $R$-symmetry breaking temperature. For $T_{\not{R}}=4kF/(\sqrt{N}\lambda \Lambda_*)$ the abundance reads 
\begin{equation}
\Omega^\text{MSSM (th)}_{3/2}h^2 \sim 0.15\,k \times \frac{16.6}{\sqrt{N}} \left(\frac{10^{10} \, \text{GeV}}{\lambda\Lambda_*} \right) \left(\frac{m_{\tilde{g}}(\mu)}{1\, \text{TeV}} \right)^2\, .
\end{equation}
We see that when three parameters $\lambda,\, \Lambda_*,\,$ and $k$ combine to the value $\lambda\Lambda_*/k \sim 10^{11}$ GeV then the gravitino can account for the dark matter of the universe. The dependence of the $\Omega^\text{MSSM (th)}_{3/2}$ on the reheating temperature is depicted in figure 3.
\begin{figure} \label{}
\centering
\begin{tabular}{cc}
{(a)} \includegraphics [scale=0.8, angle=0]{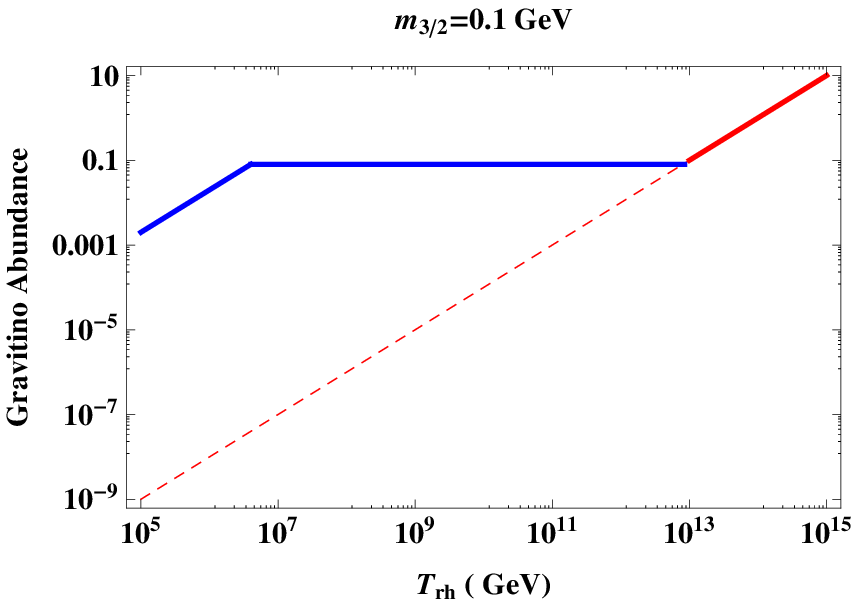}  &
{(b)} \includegraphics [scale=0.8, angle=0] {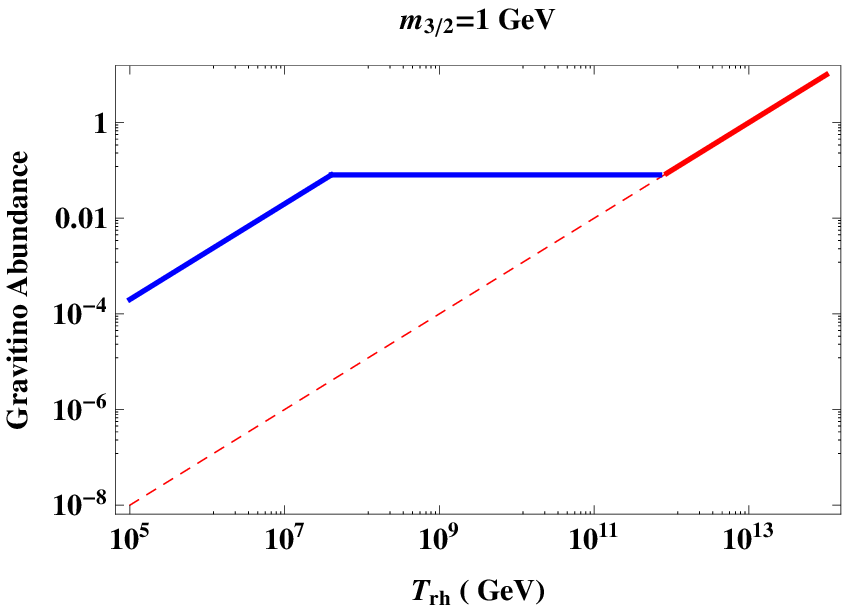}  \\
\end{tabular}
\caption{The gravitino abundance with respect to the reheating temperature for masses $m_{3/2}=0.1$ (left panel) and 1 GeV (right panel) considering \itshape only \normalfont the scatterings from the MSSM sector.  The red dashed line corresponds to the abundnance of the helicity $\pm3/2$ gravitino component and the blue line to the $\pm 1/2$ component. The hidden sector parameters were chosen so as the gravitino abundance to be $\Omega_{3/2} h^2=0.1$ for reheating temperatures larger than $T_\text{rh}$ and smaller than $10^{12} m^{-1}_{3/2} \text{ GeV}^2$. The plateau in the panels is due to the $R$-symmetry restoration i.e. due to the suppression of the helicity $\pm 1/2$ component.
}
\end{figure}

\subsubsection{Messenger sector}
Despite the fact that the $R$-symmetry restoration essentially decouples the helicity $\pm1/2$ gravitino mode from the MSSM thermal plasma it signals  the thermalization of the messenger sector. Thermalized messenger fields can efficiently exchange energy with the goldstinos.
Indeed, from the single goldstino production eq. (\ref{crossME}) we find that for 
\begin{equation} \label{goldeq}
\Gamma_{G\phi}= \left\langle n\sigma v\right\rangle =\left\langle 6 \xi  \lambda^2 k^2 T^3/s \right\rangle = 6 \xi \lambda^2 k^2 T> H(T) 
\end{equation}
the goldstinos can be also singly destroyed and hence, a thermal equilibrium for the  the helicity $\pm1/2$ gravitinos with the thermalized messengers can be achieved. This is a case much similar to the axions that can obtain a thermal abundance due to the pion-axion conversion processes. In the above relation the $H$ is the Hubble expansion scale $H(T)= 0.33 g^{1/2}_* T^2/M_\text{Pl}$ and $g_*(T)$ the effective number of degrees of freedom; assuming a pair of messenger multiplets in the ${\bold 5}+\bar{{\bold 5}}$ representation of the SU(5) and the MSSM in thermal equilibrium then it is $g_*(T\gtrsim M_\text{mess})\simeq 270$. Plugging in numbers we find that the goldstinos acquire a thermal equilibrium distribution for coupling values
\begin{equation} \label{mgeq}
\left. \lambda \, k > \sqrt{ \frac{T}{M_\text{Pl}}} \, \right|_{T=T_\text{mess}}\simeq 10^{-5} \left( \frac{T_\text{mess}}{10^8 \text{GeV}}\right)^{1/2}
\end{equation}
where a coefficient of order ${\cal O}(1)$ has been neglected.  The $T_\text{mess}$ is the temperature that verifies the equation $T= M_\text{mess}(T)\equiv \lambda X(T)$. At the temperature $T_\text{mess}$ the Boltzmann suppression of the messenger fields number density initiates which results, on the one hand, on the termination of the goldstino production from scatterings with the messengers and, on the other, on the exponential suppression of the induced thermal mass  for the spurion $X$ scalar field. The $T_\text{mess}$ does not coincide with the  zero temperature messenger mass $M_\text{mess}=\lambda X_0$. Actually, it is always $T_\text{mess} \leq M_\text{mess}$. When it is $T_\text{mess}<T_{\not{R}}$ it is a good approximation to say that $T_\text{mess}\simeq M_\text{mess}$
because at the temperature  $T_{\not{R}}$ the spurion vev has approached the $R$-violating $X_0$ value, $\left\langle X(T_{\not{R}})\right\rangle=X_0/2$.
In order to simplify the analysis and formuli we will use the $M_\text{mess}$ instead of the $T_\text{mess}$ 
 for the temperature that signals the transition of messengers to the non-relativistic regime. Actually, the final gravitino abundance is sensitive to initial temperatures when no dilution by the spurion  takes place and this happens when the approximation $T_\text{mess}\simeq M_\text{mess}$ is valid.

Coming back to the abundance of the goldstinos we see that when the condition ($\ref{mgeq}$) is true the goldstinos number density is the equilibrium one. At temperatures $T\sim M_\text{mess}$ the goldstinos decouple and the asymptotic value of the yield, $Y_{3/2,\,\infty}$, is the equilibrium value at freeze out \cite{Kolb:1990vq} $Y_{3/2,\,\infty} =Y^\text{eq}_{3/2}(T\sim M_\text{mess})  =  0.278 {g_\text{eff}}/{g_*(T)}$
where $g_\text{eff}=2\times (3/4)$ and the yield $Y_{3/2}$ is defined by the ratio of the gravitino number density $n_{3/2}$ relative to the entropy density $s$. 
In terms of energy densities, the gravitino thermal relic density over the critical one reads
\begin{equation} \label{Oth}
\left. \Omega^{\text{eq}}_{3/2}h^2 \simeq {\cal O} (5)\times 10^{5} \left( \frac{m_{3/2}}{\text{GeV}} \right) \left( \frac{270}{g_*(T)} \right)\right|_{T=M_\text{mess}} .
\end{equation} 
We stress that at $T<T_{\not{R}}$ the goldstino interactions with the MSSM gauge supermultiplets are turned on and therefore the goldstinos may stay thermalized. If this is the case then the goldstino freeze out temperature with the MSSM degrees of freedom can be smaller than the $M_\text{mess}$ value. However, for gravitino masses above the ${\cal O}$(MeV) scale the freeze out temperature usually cannot be less than $M_\text{mess}$.

Unless the condition (\ref{mgeq}) is true the goldstinos do not equilibrate with the messenger fields. Nevertheless, the messenger fields can generate a significant gravitino population. The abundance of decoupled goldstinos from scatterings with the thermalized messengers  peaks at $T\simeq M_\text{mess}$ and was given in ref. \cite{Choi:1999xm}. We write the expression for the yield and the relic abundance in a form that manifests the dependence on the coupling $\lambda$, the parameter $k$ and the messenger mass $M_\text{mess}$ in transparent way rather than the gaugino mass whose value is irrelevant in this kind of processes (irrelevant in the sense that there is no the $m^2_\lambda/m^2_{3/2}$ coupling strength coefficient here): 
\begin{equation}\label{scayield}
Y^\text{mess(sc)}_{3/2}=3.1 \times 10^{-4}\left(\frac{10^5}{M_\text{mess}} \right)\left(\frac{\lambda k}{10^{-6}} \right)^2 \left( \frac{270}{g_*}\right)^{3/2}
\end{equation}
and $\Omega^{\text{mess(sc)}}_{3/2}h^2 \simeq \,2.8 \times 10^8 Y^\text{mess(sc)}_{3/2} {m_{3/2}} {\text{GeV}^{-1}} $.
For $\bar{\Lambda}\sim 1.5 \times 10^5$ GeV the relic density takes the simple form $\Omega^{\text{mess(sc)}}_{3/2}h^2 \simeq 0.3 \times10^9 \lambda k$. Equally, utilizing the relation (\ref{barL}) 
the abundance can be recast in the more familiar-looking form:
\begin{equation}
\Omega^{\text{mess(sc)}}_{3/2}h^2 \sim \left(\frac{\text{GeV}}{m_{3/2}} \right) \left(\frac{ M_\text{mess} }{10^5\, \text{GeV}} \right) \left(\frac{m_\lambda}{\text{TeV}}\right)^2.
\end{equation}
These expressions give the goldstino abundance from the scatterings of the messengers in the plasma. Some technical details for the calculation of the $Y^\text{mess (sc)}_{3/2}$ can be found in the appendix B.

\subsection{Non-thermal production of gravitinos and dilution}

Apart from the gravitino emission in scatterings of thermalized supersymmetric particles there is also a second source that contributes to the final gravitino abundance: the decay of the plasma ingredients plus decoupled species that have potentially dominated the energy density of the universe. Hereafter, we are going to examine and present the gravitino production from the decays of the MSSM particles, the messengers and the spurion.

\subsubsection{MSSM decays}
The contribution to the $\Omega_{3/2}$ from the Standard Model sparticles decays does not depend on the reheating temperature. It peaks at temperatures $T\sim \tilde{m}$ where $\tilde{m}$, a soft mass. Utilizing the decay width into gravitinos (\ref{ssmd}) the relic density of gravitinos from MSSM decays is found to be \cite{Drees:2004jm}
\begin{equation}
  \Omega^{\text{MSSM(dec)}}_{3/2} h^2 \simeq 
3.4 \times 10^{-5} \left(\frac{\text{GeV}}{m_{3/2}} \right) \left(\frac{228}{g_*(\tilde{m})}\right)^{3/2}
\left(\frac{\tilde{m}}{\text{TeV}}\right)^3.
\end{equation}

\subsubsection{Messenger decays}
There is no basic model building reason for the messenger particles of gauge mediation supersymmetry breaking models to be unstable. Actually, if the messenger fields have no direct couplings with the MSSM sector then the theory conserves a global messenger quantum number that can render the lightest messenger particle stable and therefore relic of the early universe \cite{Dimopoulos:1996gy}. The stable messenger can have an acceptable thermal relic population only under particular assumptions a fact that makes the stable messenger scenario not attractive. Even if the lightest messenger is unstable it may have an impact on the cosmological evolution by the production of late entropy. The decay of the heavier messenger particles can also generate an important number of non-thermal goldstinos. Actually, the decay channels of the messenger particles influence the final gravitino abundance.

The decay width of messenger particles to goldstinos is dominated by the superpotential coupling (\ref{mess2}) and takes the value $\Gamma_\lambda = k^4 \lambda^4 F^2 /(4\pi M^3_\text{mess})$. Also in this case the goldstino production from decays of the messengers
 peaks at temperatures of order of the $M_\text{mess}$.

We recall that the messenger sector consists of a $N$ pairs of chiral supermultiplets $\phi+\bar{\phi}$ and each pair describes a Dirac fermion with mass $M\equiv M_\text{mess}$ and two complex scalar particles with mass squared $M^2\pm \lambda F$. Taking into account the SU(2) D-terms the mass of the scalars within the same isospin multiplet are further split. We denote the ligthest messenger as $\phi_{LM}$. The scalar $\phi_{LM}$ particle is lighter than its fermionic partner hence it cannot decay via the superpotential interaction (\ref{mess2}) to a messenger fermion and a goldstino. If the messenger sector has a conserved quantum number corresponding to an accidental global symmetry the lightest messenger will be a stable scalar particle.
According to the analysis of \cite{Dimopoulos:1996gy}, if the lightest messenger sits in representations of SU(5) it is either the neutral or the charged component of a weak doublet in the case of $\bold{5}+\bar{\bold{5}}$ or a weak singlet with a unit of electric charge in the case of $\bold{10}+\bar{\bold{10}}$. The $\phi_{LM}$ can be also an SU(3)$\times$ SU(2)$\times$ U(1) singlet in the case of $\bold{16}+\bar{\bold{16}}$ representations of SO(10). For the first two cases the lightest messenger decouples from the thermal plasma when it is non-relativistic and the yield is roughly $Y_{\phi_{LM}} \simeq 3.7 \times 10^{-10} M_{\phi_{LM}}/10^6 \text{GeV}$ which gives the abundance $\Omega_{\phi_{LM}} \simeq 2.8\times10^8 Y_{\phi_{LM}} M_{\phi_{LM}}/\text{GeV}$ i.e.
\begin{equation}
\Omega_{\phi_{LM}} h^2 \simeq  10^{5} \left( \frac{M_{\phi_{LM}}}{10^6\,\text{GeV}}\right)^2.
\end{equation}
If the $\phi_{LM}$ is a Standard Model singlet it decouples when it is still relativistic with yield ${\cal O}(10^{-3})$ and its relic density is 
$\Omega_{\phi_{LM}} h^2 \simeq 7\times 10^{11} M_\text{mess}/(10^6\,\text{GeV})$ given that the reheating temperature was about or higher than the GUT scale where, actually, the thermal equilibrium is doubtful. However, there is a crucial point here that has to be taken into account: the supersymmetry preserving vacuum is the attractive minimum of the finite temperature effective potential \cite{Dalianis:2010pq} since the temperature corrections for the SM singlet $\phi_{LM}$ are negligible and cannot compensate the tachyonic directions during the reheating phase. For this reason, we are not going to consider the Standard Model singlet $\phi_{LM}$ case altogether\footnote{We note, however, that superpotential couplings of the form (\ref{Winter}) with sufficiently large Yukawa couplings could block the evolution of the system of fields towards the supersymmetric minimum.}.

The lightest messenger can be a dark matter component without overclosing the universe only under particular conditions hence, messenger number violating interactions that allow the lightest messenger to decay are necessary. An extended study of the messenger couplings in a cosmological context has been carried out in ref. \cite{Jedamzik:2005ir}. Generally, superpotential couplings between the messenger and the matter sector are not problem-free. Although the SM gauge symmetries allow a large number of superpotential couplings there are constraints from the proton decay limits, from negative contributions to squared sfermions masses generated at one loop level and from bounds on FCNC processes. Nevertheless, extra marginal messenger interactions seem to be significant from a particle physics phenomenological point of view as well. Indeed, the difficulty of the gauge intractions in generating appropriate contributions to Higgs sector soft parameters, given that $m_h\approx 126$ GeV, suggests that a complete theory of supersymmetry breaking at the weak scale should include both gauge mediation and additional coupling to MSSM and in particular the Higgs sector. In ref. \cite{Evans:2013kxa} a recent classification of all possible such interactions can be found. Here we give the messenger-matter couplings which were firstly introduced in the ref. \cite{Dine:1996xk} and have the form 
\begin{equation} \label{Winter}
W \supset y^i_l H_D \phi_2 \bar{e}_i + y^i_q H_D Q_i \bar{\phi}_3\, ,
\end{equation}
where $\phi_2$ and $\bar{\phi}_3$ are lepton and quark like messengers, $H_D$, $\bar{e}$, $Q_i$ are respectively SM down-type Higgs, right-handed lepton and quark doublet and the $y^i$  are Yukawa couplings with family index $i$. 
Tachyonic sfermion masses are avoided for $\sum |y|^2<10^{-3}$ \cite{Drees:2004jm}. Flavour changing neutral currents lead to even stronger constraints. Assuming conservatively that $y^1_l \neq 0$ and zero Yukawa couplings with the second and third families then there is the limit on this Yukawa coupling $y^1_l \lesssim (M_\text{mess}/10^8 \text{GeV})$ \cite{Jedamzik:2005ir}. The superpotential couplings (\ref{Winter}) induce a decay width 
\begin{equation} \label{Gw}
\Gamma_y=\frac{y^2 M_\text{mess}}{8\pi} \,.
\end{equation}
Unless the Yukawa couplings $y$ are extremely small the decay of the messengers  takes place fast without entropy production i.e. before the domination of the energy density of the universe by the messenger fields.  Actually, the decay rate (\ref{Gw}) can be larger not only from the Hubble rate at the temperature $T\sim M_\text{mess}$ but also from the pair creation rate of the messenger particles themselves. Indeed, when $\Gamma_{y} > \Gamma_\text{int} = \left\langle n \sigma v \right\rangle$ where $\sigma \sim \alpha^2/T^2$ the messengers particles decay more rapidly than the rate produced by $2 \leftrightarrow 2$ scatterings in the thermal plasma. This simply means that for $T>M_\text{mess}$ the messenger particles might have an equilibrium number density due to the inverse decays of MSSM particles to messengers, which occur with inverse decay rate $\Gamma_{\text{I}y} \simeq \Gamma_\text{y}$, than due to the two-to-two scatterings. It is worth to notice that if $\Gamma_{\text{I}y} \ll \Gamma_{y}$, i.e. T (or CP) invariance was violated, then it might have been possible to significantly suppress the messenger number density in the plasma and hence, the subsequent goldstino production. However, there is no any (at least profound) physical reason for such a suppresion to happen.
The inverse decay rate of the messengers has the following temperature dependence \cite{Kolb:1990vq}
$$
\Gamma_{\text{I}y}(T) \simeq \Gamma_y
\begin{cases} 
1 &  T \gtrsim M_\text{mess} \\ 
(M_\text{mess}/T)^{3/2}e^{-M_\text{mess}/T} & T\lesssim M_\text{mess} 
\end{cases}
$$  

Nevertheless, another kind of interaction might exist and, unless the superpotential couplings (\ref{Winter}) are present, can induce the decay of the lightest messenger.
 If this interaction is weak enough then it can cause a late entropy production.  One can assume a messenger-matter mixing due to a correction in the superpotential $\delta W \simeq (\left\langle W \right\rangle/M^2_\text{Pl})\bold{5}_M \bold{\bar{5}}_F$ where the subscripts $M$ and $F$ denote the multiplet that belongs to the messenger sector and MSSM matter respectively and $\left\langle W \right\rangle \simeq m_{3/2}M^2_\text{Pl}$. It was proposed in ref. \cite{Fujii:2002fv} and it provides a decay channel of $\phi_{LM}$ into a Standard Model lepton and a gaugino with decay rate, $\Gamma_m$,  estimated to be 
\begin{equation} \label{mix}
\Gamma_m = \, \frac{g^2_2}{16\pi} \frac{m^2_{3/2}}{M_{\text{mess}}}\,.
\end{equation}
We have assumed small mass splittings between the messenger particles, i.e. $M^2_\text{mess} \gg \lambda F$, and for order of magnitude estimations we use approximately the $M_\text{mess}$ instead of the $M_{\phi_{LM}}$ for the mass of the lightest messenger.
\\
\\
\itshape{Decay widths hierarchy and goldstino yield}\normalfont
\\
The amplitudes of the various decay rates of the messenger fields are essential for the estimation of the final gravitino abundance. As we will argue, the $\Gamma_y$, $\Gamma_m$ and $\Gamma_\lambda$ values not only determine whether late entropy production takes place but, also, their ratio controls the goldstino yield from the messenger decays. Comparing the messenger-matter decay rates (\ref{Gw}) and (\ref{mix}) with the messengers-goldstino rate (\ref{Gmess}) one finds  that for $\lambda k \ll 1$ it is $\Gamma_y \gg \Gamma_{\lambda}$ and $\Gamma_m \gg \Gamma_\lambda$, as illustrated in figure 4. In particular, taking the Yukawa coupling $y=0.1 M_\text{mess}/10^8$ GeV such that the flavour changing neutral current constraints are satisfied then
\begin{equation} \label{lambday}
\frac{\Gamma_\lambda}{ \Gamma_y} \simeq 2\times 10^{-12} \left(\frac{\lambda k}{10^{-6}} \right)^2\left( \frac{\bar{\Lambda}}{10^5\, \text{GeV}}\right)^2 \left(\frac{10^7 \text{GeV}}{M_\text{mess}} \right)^4\,
\end{equation}
and
\begin{equation} \label{lambdam}
\frac{\Gamma_\lambda}{\Gamma_m} \simeq 1.4 \left(\frac{\lambda k}{10^{-6}} \right)^4 \left(\frac{10^7\, \text{GeV}}{M_\text{mess}} \right)^2\,.
\end{equation}
In other words, for $\lambda k\ll 1$, the messenger particles are much more probable to decay to MSSM particles rather than to goldstino fields. The total decay rate of the messenger fields is $\Gamma_\text{tot}=\Gamma_\lambda+\Gamma_y+\Gamma_m$, therefore the goldstino yield from the messenger decays is 
\begin{equation} \label{dyield}
Y^\text{mess(dec)}_{3/2} \sim B_{3/2} N_{\phi\bar{\phi}} Y^\text{eq}_\text{mess}(T\sim M_\text{mess})
\end{equation}
where $B_{3/2}=\Gamma_\lambda/\Gamma_\text{tot} < 1$ and $N_{\phi\bar{\phi}}$ the messenger degrees of freedom that decay to goldstino and a lighter messenger superpartner. The goldstino yield from the decays is dominated by temperatures $T\sim M_\text{mess}$. Actually, the $\Gamma_\text{tot}$ is much larger than the Hubble rate when messengers become non-relativistic. 

In the figure 4 the scaling of the three different decay rates is shown for different messenger mass scales. It is manifest that for $\lambda k<10^{-6}$ it is $\Gamma_y+\Gamma_m \gg \Gamma_\lambda$ and the $Y^{\text{mess(dec)}}_{3/2}$, according the expression (\ref{dyield}) is smaller than the goldstino yield from scatterings  (\ref{scayield}). Taking into consideration the relation (\ref{mgeq}), i.e. the fact that the goldstinos acquire a thermal abundnace for $\lambda k >10^{-5} (M_\text{mess}/10^8\text{GeV})^{1/2}$, we conclude that we always have
\begin{equation} \label{domi}
Y^\text{mess(dec)}_{3/2} < Y^\text{mess(sc)}_{3/2} \leq Y^\text{eq}_{3/2}
\end{equation}
regardless the messenger scale (see also appendix B).

Due to the fact that we explicitly consider messenger-MSSM interactions our result is different than that of ref. \cite{Choi:1999xm} and \cite{Jedamzik:2005ir} which find that the goldstino yield from the decays of the messenger fields with mass $M_\text{mess}\lesssim 10^6 \,\text{GeV}$ dominates over the goldstino yield from scatterings. Therefore here the goldstinos are produced more efficiently from the scatterings rather than the decay of the thermalized messengers and we will consider the $Y^\text{mess(dec)}_{3/2}$ as a small correction to the final gravitino yield. The $Y^\text{mess(dec)}_{3/2}$ is important only when the messengers dominate the energy density of the universe before their final decay.
\begin{figure} \label{dw}
\centering
\begin{tabular}{ccc}
{(a)} \includegraphics [scale=0.53, angle=0]{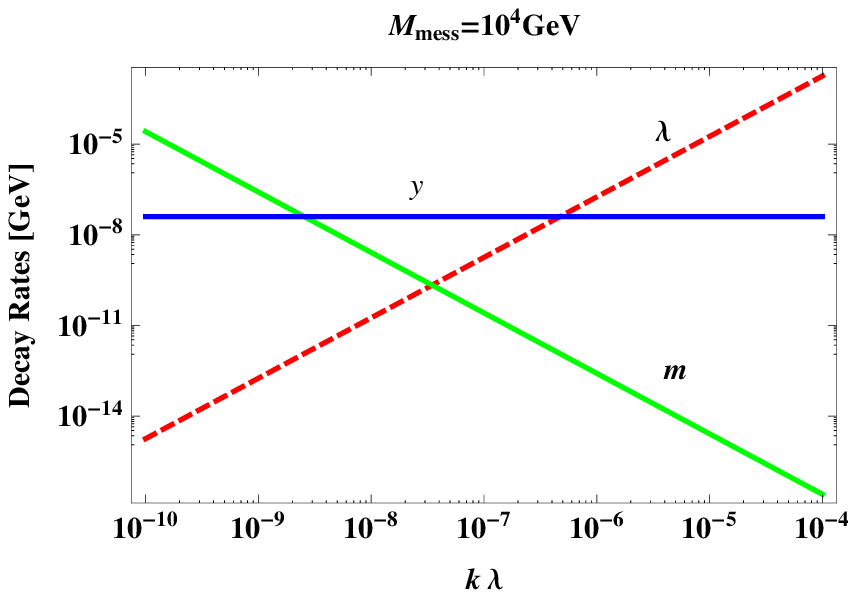}  &
{(b)} \includegraphics [scale=0.53, angle=0]{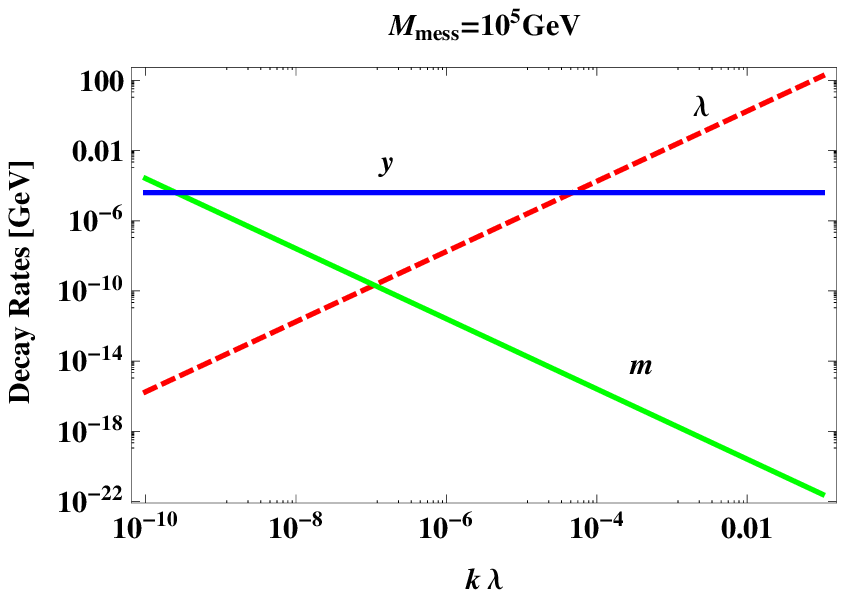}  &
{(c)} \includegraphics [scale=0.53, angle=0] {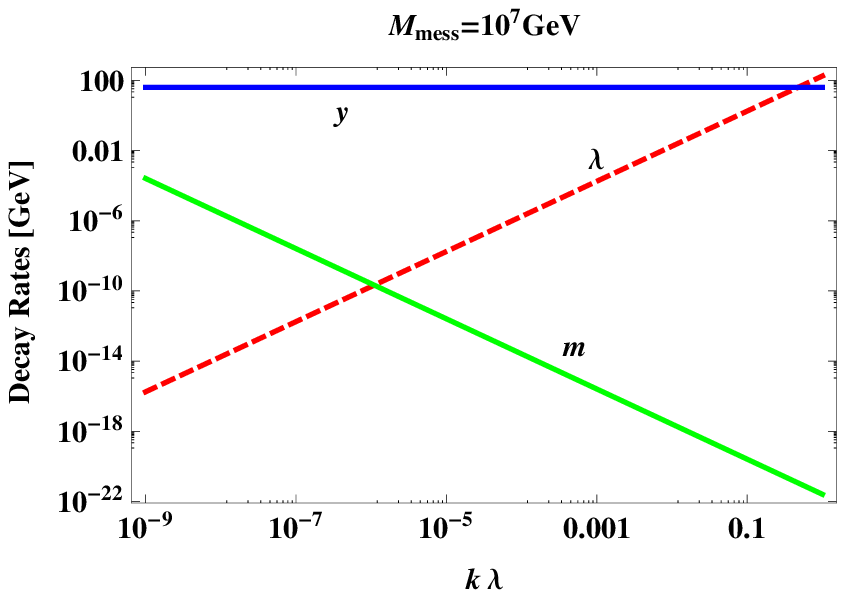}  \\
\end{tabular}
\caption{The three decay rates of the messenger particles to goldstinos (red dashed line -$\lambda$) and to MSSM particles (continuous lines, blue -$y$ for superpotential Yukawa couplings (\ref{Winter}) and green -$m$ for the correction (\ref{mix})). The three panels correspond to different messenger masses, $M_\text{mess}=10^4\,,10^5\,, $ and $10^7$ GeV (from left to right). It is manifest that for $k\lambda \ll 1$ the messengers dominantly decay to MSSM degrees of freedom.} 
\end{figure}

\subsubsection{Dilution from lightest messenger decay}

Let us assume, here, that there are no direct messenger-matter couplings in the superpotential i.e. $\Gamma_y=0$. However, in order the messenger sector not to be totally isolated from the observable, (this would imply that the lightest messenger particle is stable) we take, in this subsection $\Gamma_y=0$ and $\Gamma_m \neq 0$. Vanishing direct messenger-matter direct mixings makes the lightest messenger $\phi_{LM}$ longlived. Unless $\phi_{LM}$ is a Standard Model singlet and $\Gamma_m$ is weak enough it will decouple from the thermal plasma at the temperature $T_f \sim M_{\phi_{LM}}/20$. Afterwards, the energy density of the $\phi_{LM}$, $\rho_{\phi_{LM}}=2\pi^2 g_*(T)T^3 M_\text{mess} Y_{\phi_{LM}} /45$, will scale like $a^{-3}$ contrary to the $a^{-4}$ scaling of the radiation, where $a$ is the scale factor of the universe. The temporary relic messengers $\phi_{LM}$ become the largest component of the Universe energy density at the temperature
\begin{equation}
T_\text{dom}=\frac{4}{3}M_\text{mess}  Y_{\phi_{LM}} \simeq 5\times 10^{-4} \left(\frac{M_\text{mess}}{10^6 \,\text{GeV}} \right)^2 \text{GeV},
\end{equation}
where, for simplicity, we used the approximation $M_\text{mess} \approx M_{\phi_{LM}}$. The $\phi_{LM}$ will decay due to the "indirect" messenger-matter mixing \cite{Fujii:2002fv} at the temperature  
\begin{equation}
T_{\phi_{LM}} \simeq 22 \,\text{GeV} \, k^{-1/2} \left( \frac{10}{g_*(T)}\right)^{1/4} \left(\frac{\bar{\Lambda}}{10^5 \text{GeV}} \right)^{1/2} \left(\frac{m_{3/2}}{\text{GeV}} \right)^{1/2}.
\end{equation}
If $T_{\phi_{LM}} \ll T_\text{dom}$ then the thermal relic $\phi_{LM}$ dominate the energy density, i.e. $\rho_\text{tot} \simeq \rho_\text{rad} +\rho_m \gg \rho_\text{rad}$ where $\rho_\text{rad}=\pi^2g_*(T)T^4/30$,  and its subsequent decay will produce significant entropy, reheating the Universe and diluting any pre-existing abundances. The entropy release is given by the dilution factor $\Delta_\text{mess}$
\begin{equation} \label{Dmess}
\Delta_\text{mess}=\frac{T_\text{dom} }{T_{\phi_{LM}}} =4.6 \times k \sqrt{\lambda} \left(\frac{M_\text{mess}}{10^8\, \text{GeV}} \right)^{3/2}\,.
\end{equation}
By definition, the dilution factor cannot be less than one. Apparently, there is no dilution from the lightest messenger decay if the messenger mass, the messenger superpotential coupling $\lambda$ and the hidden sector parameter $k$ are small enough. Replacing the $m_{3/2}$ from the relation (\ref{barL}) we find that there is no dilution from messenger decays, i.e. $T_\text{dom}<T_{\phi_{LM}}$, when
\begin{equation} \label{nodil}
\lambda k^2 < 38 \left(\frac{10^6 \text{GeV}}{M_\text{mess}}\right)^3 \left( \frac{\bar{\Lambda}}{10^5\text{GeV}} \right).
\end{equation}
Figures 5-7 illustrate this condition for different messenger scales. The absence of dilution may not be problematic because, firstly, the goldstino abundance from scatterings in the plasma, given by the expression (\ref{scayield}), has a dependence $Y^{\text{mess(sc)}}_{3/2} \propto \lambda^2 k^2$ and it can be small enough. Secondly, the spurion may be also a source of late entropy production.
For $\lambda k >10^{-5} (M_\text{mess}/10^8\text{GeV})^{1/2}$ the goldstinos get thermalized and after the late entropy production the gravitino relic density obtains the value
\begin{equation}
\frac{\Omega^\text{eq}_{3/2}}{\Delta_\text{mess}} = 0.12 \frac{1}{\lambda^{3/2}k^2} \left(\frac{10^{8}\,\text{GeV}}{M_\text{mess}} \right)^{1/2}\,.
\end{equation}
The dilution of the gravitinos from the lightest messenger particle decay induced by the superpotential coupling $\delta W \simeq (\left\langle W \right\rangle/M^2_{P})\bold{5}_M \bold{\bar{5}}_F$ was proposed in ref. \cite{Fujii:2002fv}. However, there is a small difference: here the goldstinos acquire a thermal abundance due to the scatterings of the messenger fields for any temperature $T>M_\text{mess}$ regardless the gravitino mass value, whereas in ref. \cite{Fujii:2002fv} the goldstinos thermalize due to MSSM fields, i.e. the messenger contribution was ignored.

\subsubsection{Spurion Decay}
A basic advantage of gauge mediation schemes is the absence of the Polonyi-like problem. The superpotential coupling $\delta W =\lambda X \phi\bar{\phi}$ decreases the lifetime of the spurion rendering it harmless for the Big Bang Nucleosynthesis (BBN) processes. Nevertheless, the spurion decays slowly and if there is a scalar $X$ condensate that dominates the energy density of the unverse then dilution and entropy production take place.

The $R$-symmetry restoration happens when the vev of the spuion, $X$, becomes vanishing. Thus in an $R$-symmetric thermal phase the spurion finds itself dispaced from its zero temperature vacuum expectation vaule $X_0$. When the temperature falls to the $T_{\not{R}}$ value the minimum of the effective potential moves to $X_0$ and the scalar field $X$ follows the minimum without sizeable oscillations only if the thermal mass for the $X$ field is not too weak \cite{Dalianis:2010pq, Fukushima:2012ra}.  Let us be more specific here. Taking the second derivative in the spurion direction of the finite temperature effective potential, the effective mass for the spurion was found to be \cite{Dalianis:2010yk}
\begin{equation}
m^2_X(T) \equiv V^T_{X\bar{X}} \simeq 4\frac{F^2}{\Lambda^2_*}+N\frac{\lambda^2T^2}{4} \,.
\end{equation}
At the temperature $T_{\not{R}}=4F/(\lambda \Lambda_*)$ the thermal correction to the spurion mass squared, $N\lambda^2 T^2/4$, becomes equal to the zero temperature mass squared. The spurion field follows the minimum which moves, roughly instantaneously, towards the zero temperature vacuum expectation value when its mass is larger than the Hubble scale. At the temperature $T_{\not{R}}$ the condition $m_X>H$ translates into
\begin{equation} \label{osci}
\lambda \gtrsim 5 \frac{T_{\not{R}}}{M_\text{Pl}} 
\end{equation}
Taking the (\ref{min-3}) model with $\delta W=c$ as a representative example for an order of magnitude estimation of the above constraint on the coupling we find
$\lambda \gtrsim 10^{-8} \left({\Lambda_*}/{10^{15}\text{GeV}} \right)^{1/2} \left({\bar{\Lambda}}/{10^5\text{GeV}} \right)^{1/2}$.
Although the constraint (\ref{osci}) is necessary it is not sufficient. One should not ignore that the thermal mass for the spurion vanishes altogether when the messengers become non-relativistic (the $X$-field "decouples"). Hence, it has to be $T_{\not{R}}>M_\text{mess}(T_{\not{R}})=\lambda X(T_{\not{R}})=\lambda b_R(T_{\not{R}})X_0=M_\text{mess}/2$ (see appendix A) which translates into the bound
\begin{equation} \label{dcp}
\lambda \lesssim \, 2\times 10^{-5} \left( \frac{10^{15}\text{GeV}}{\Lambda_*}\right)^{1/2} \left(\frac{\bar{\Lambda}}{10^5\text{GeV}}\right)^{1/2} \,.
\end{equation}
Unless constraints (\ref{osci}) and (\ref{dcp}) are both satisfied the energy stored in the spurion oscillations dominate the energy density of the universe and its subsequent decay will dilute the pre-existing abundances, see figures 5-7. The dilution magnitude, $\Delta_\text{spur}$ was calculated in the ref. \cite{Fukushima:2012ra} and was found to vary from $\Delta_\text{spur}=1$ to $\Delta_\text{spur}=10^9$. The gravitino yield generated by the spurion decay, $Y^{\text{spur}}_{3/2}$,  defined by the ratio $n_{3/2}/s$ where $s$ the entropy density produced can be evaluated as
\begin{equation}
Y^{\text{spur}}_{3/2} \equiv \frac{n_{3/2}}{s}= \frac{3}{2}B^\text{spur}_{3/2}\frac{T_{\text{spur}}}{m_X}\,.
\end{equation}
The $T_\text{spur}$ stands for the spurion decay temperature and $B^\text{spur}_{3/2}$ for the branching ratio of the two gravitino decay mode. They are calculated in ref. \cite{Ibe:2006rc, Hamaguchi:2009hy} to be $T_\text{spur} = {\cal O}(0.1-10)$ GeV and $B^\text{spur}_{3/2}={\cal O}(10^{-10}-10^{-6})$. Summing up, the gravitino relic density after the reheating of the universe from the spurion decay reads
\begin{equation}
\Omega_{3/2} =\frac{\Omega^{\text{th}}_{3/2}}{\Delta_\text{spur}}+m_{3/2}Y^\text{spur}_{3/2} \frac{s_0}{\rho_{\text{cr},0}}
\end{equation}
where  $\Omega^{\text{th}}_{3/2}=\Omega^{\text{eq}}_{3/2}$ for $\lambda k>(M_\text{mess}/M_\text{Pl})^{1/2}$  or $\Omega^{\text{th}}_{3/2}=\Omega^{\text{mess(sc)}}_{3/2}+\Omega^{\text{MSSM}}_{3/2}+\Omega^{\text{grav}}_{3/2}$ otherwise. The $\rho_{\text{cr},0}/s_0\simeq 3.6 \times 10^{-9}h^{-2}$ GeV is the critical density divided by the entropy density at present. There are parameter regions where the dark matter is mostly non-thermally or a mixture of thermally (diluted) and non-thermally produced gravitinos \cite{Fukushima:2012ra}.

\subsection{Total abundance of the thermally and non-thermally produced gravitinos}

Here we sum up the gravitino relic abundance from the MSSM and the messenger sector. The final result depends on the reheating temperature and values of the coupling $\lambda$ and the hidden sector parameter $k$.

\begin{description}

\item[A)] \itshape Thermally produced helicity $\pm3/2$ gravitinos \normalfont

The helicity $\pm 3/2$ component interacts gravitationally with the thermal plasma. It can be produced mainly from $2 \rightarrow 2$ scatterings with thermalized messenger and MSSM particles. The cross section is independent of the temperature (apart from the running of the gauge couplings) and hence the relic abundance has a linear dependence on the temperature. This contribution is always present:
\begin{equation}
\Omega^\text{grav}_{3/2}h^2 \simeq 0.1 \left(\frac{m_{3/2}}{\text{GeV}}\right) \left( \frac{T_\text{rh}}{10^{12}\text{GeV}} \right)
\end{equation}

\item[B)] \itshape Thermally plus non-thermally produced helicity $\pm1/2$ gravitinos  from MSSM sector\normalfont
\begin{enumerate}
\item $T_\text{rh}>\text{max} \{ M_\text{mess}, T_{\not{R}} \}$
\begin{equation} \label{ORs}
\Omega^\text{MSSM (sc)}_{3/2}h^2 \sim 0.15 \times \frac{16.6}{\sqrt{N}} \left(\frac{10^{10} \, \text{GeV}}{\lambda\Lambda_*/k} \right) \left(\frac{m_{\tilde{g}}(\mu)}{1\, \text{TeV}} \right)^2\, 
\end{equation}

\item $\tilde{m}/10\lesssim T_\text{rh}<\text{max} \{ M_\text{mess}, T_{\not{R}} \}$
\begin{equation}  \label{OMSSM}
\Omega^\text{MSSM (sc)}_{3/2}h^2 \simeq 0.2 \left(\frac{ T_{rh}}{10^8\,\text{GeV}} \right) \left(\frac{\text{GeV}}{m_{3/2}} \right) \left(\frac{m_{\tilde{g}}(\mu)}{1\, \text{TeV}} \right)^2\, 
\end{equation}

\item $T_\text{rh}>\tilde{m}/10$
\begin{equation}
  \Omega^{\text{MSSM(dec)}}_{3/2} h^2 \simeq 
3.4 \times 10^{-5} \left(\frac{\text{GeV}}{m_{3/2}} \right) \left(\frac{228}{g_*(\tilde{m})}\right)^{3/2}
\left(\frac{\tilde{m}}{\text{TeV}}\right)^3.
\end{equation}

\end{enumerate}

\item[C)] \itshape Thermally plus non-thermally produced helicity $\pm 1/2$ gravitinos from messengers  \normalfont :\\ i.e. $T_\text{rh}>M_\text{mess}$
\begin{enumerate}
	
	\item Thermal goldstinos, 	$\ \lambda k >(M_\text{mess}/M_\text{Pl})^{1/2}$:

\begin{equation}
\Omega^\text{eq}_{3/2}h^2  \, \sim \, {\cal O}(5)\times 10^5 \left(\frac{m_{3/2}}{\text{GeV}} \right) \left( \frac{270}{g_*(T)} \right) 
\end{equation}	

\item Thermally produced goldstinos (from scatterings),  $\lambda k <(M_\text{mess}/M_\text{Pl})^{1/2}$:
	
\begin{equation}
\Omega^\text{mess(sc)}_{3/2} h^2  \, \sim \, 3.7\, \lambda^2 k^2 \left(\frac{M_\text{Pl}}{M_\text{mess}}\right) \left( \frac{m_{3/2}}{\text{GeV}}\right) \left( \frac{270}{g_*(T)} \right) 
\end{equation}	
\end{enumerate}
\end{description}
\itshape Summing up \normalfont\\
\begin{itemize}
\item \bfseries Total gravitino abundance \normalfont (when spurion's oscillations are adiabatically suppressed):
\end{itemize}
\begin{equation}
\Omega_{3/2}(T_\text{rh})=\text{min}\left\{\Omega^\text{mess}_{3/2}+\Omega^\text{MSSM}_{3/2}+\Omega^\text{grav}_{3/2}, \,\,\,\Omega^\text{eq}_{3/2} \right\}
\end{equation}
\\
when no dilution from the lightest messenger decay takes place. The gravitino abundance is diluted by the lightest messenger decay, i.e. $\Omega_{3/2} \rightarrow  \Omega_{3/2}/\Delta_\text{mess}$, only under the special conditions: $\Gamma_y = 0$	\,  and  \,  $\lambda k^2 > 38 \,  (10^6\text{GeV}/M_\text{mess})^3 \left({\bar{\Lambda}}/{10^5\text{GeV}} \right)$.
\\
\begin{description}
\item[D)] \itshape Non thermally produced goldstinos from the spurion decay	and spurion dilution \normalfont
\end{description}
$$
\begin{Bmatrix} 
T_\text{rh}>\text{max} \{ M_\text{mess}, T_{\not{R}} \} & \\ 
 \lambda < 5\,T_{\not{R}}/ M_\text{Pl}$\,\, or \,\,$ M_\text{mess} > 2T_{\not{R}} &
\end{Bmatrix}  \leftrightarrow \,\text{spurion oscillations are undamped}
$$
Hence,
\begin{itemize}
\item \bfseries Total gravitino abundance \normalfont (when spurion's oscillations dominate the energy density):
\end{itemize}
\begin{equation} \label{OSpu}
\Omega_{3/2}(T_\text{rh}) =\frac{\text{min}\left\{\Omega^\text{mess}_{3/2}+\Omega^\text{MSSM}_{3/2}+\Omega^\text{grav}_{3/2}, \,\,\,\Omega^\text{eq}_{3/2} \right\}}{\Delta_\text{spur}}+ 2.8 \times 10^8 \, Y^\text{spur}_{3/2} \left(\frac{m_{3/2}}{\text{GeV}}\right)\, h^{-2}
\end{equation}
\\
If there is $\Delta_\text{mess}>1$ then the final dilution factor is determined solely by the $\Delta_\text{spurion}$ since the spurion decays slower  than the lightest messenger (though the $T_{\phi_{LM}}$ is more model dependent). Actually, it happens that the parameter regions that implement $\Delta_\text{mess}>1$ and $\Delta_\text{spurion}>1$ largely overlap (figure 7). Thus, except for special cases, the potential dilution effects of the lightest messenger decay can be neglected.
\\
\\
\\
\begin{figure} \label{k-l-1}
\centering
\begin{tabular}{cc}
{(a)} \includegraphics [scale=0.8, angle=0]{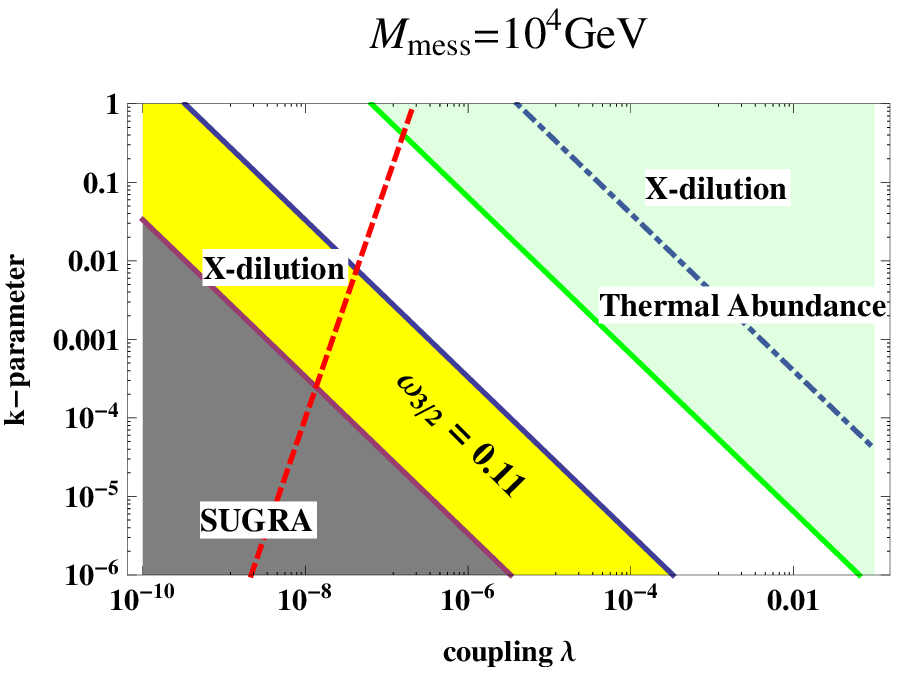}  &
{(b)} \includegraphics [scale=0.8, angle=0] {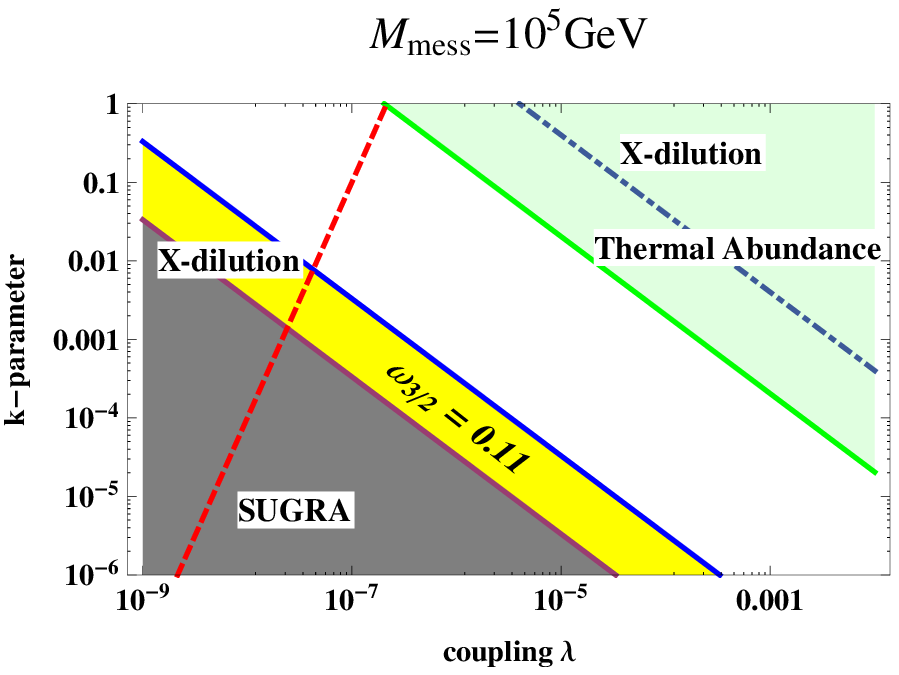}  \\
\end{tabular}
\caption{Contour plots of the  helicity $\pm1/2$ gravitino abundance for different values of the superpotential coupling $\lambda$, the parameter $k\equiv F_X/F$ and low messenger mass $M_\text{mess}$. The gray region (lower left corner) corresponds to gravitino mass $m_{3/2} \geq 100$ GeV where gravity effects play a dominant role. In the rest contour area it is $m_{3/2} < 100$ GeV. The light coloured region above the green line (upper right corner) corresponds to gravitinos with thermal abundance. In the white part of the contour the gravitino is not thermalized,  albeit it is overabundant due to messengers scatterings. The yellow strip corresponds to $\lambda \cdot k$ values that render the gravitino production from messenger fields inefficient thus, the gravitino production is dominated by the MSSM gauge superfields; the gravitino relic density $\omega_{3/2} \equiv \Omega_{3/2}h^2=0.11$ can be realized. In the two areas between the upper corners of the axes and the dashed lines dilution by the spurion decay takes place.}
\end{figure}
\begin{figure} \label{k-l-2}
\centering
\begin{tabular}{cc}
{(c)} \includegraphics [scale=0.8, angle=0]{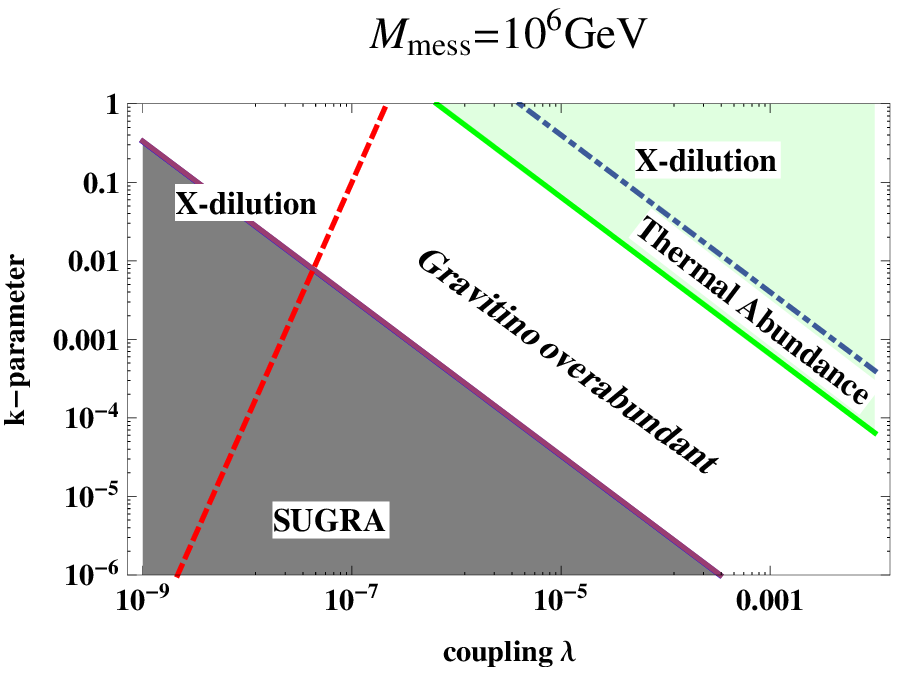}  &
{(d)} \includegraphics [scale=0.8, angle=0] {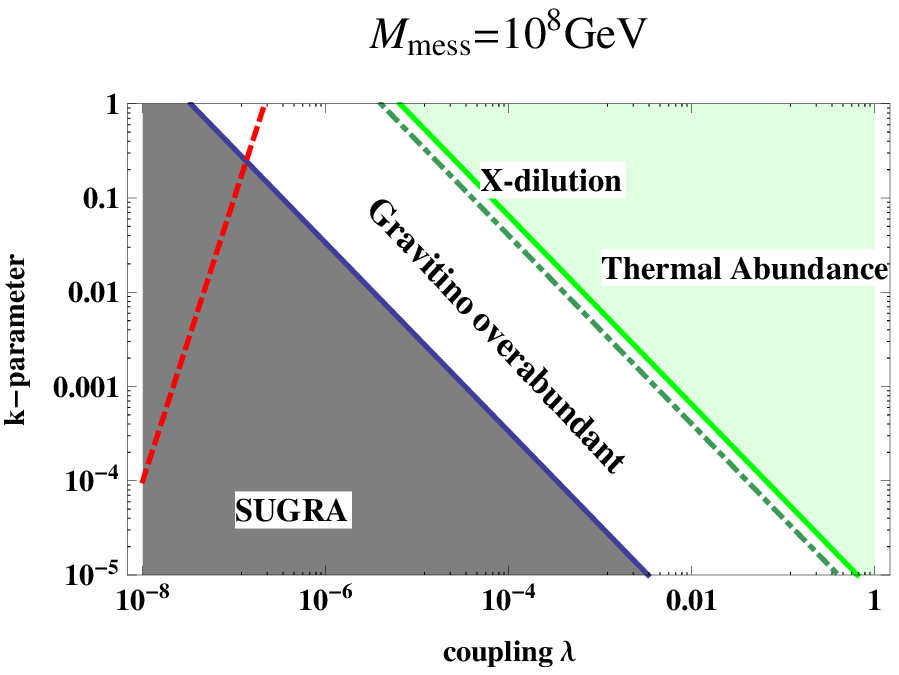}  \\
\end{tabular}
\caption{The same with figure 5 contour plots for intermediate messenger mass $M_\text{mess}$. There is no (yellow) area in this contour where $\Omega_{3/2}h^2 \leq 0.11$ is possible without late entropy production . The overabundant gravitinos can de diluted by the spurion decay which takes place in the two areas between the upper corners of the axes and the dashed lines. The correct magnitude of the dilution takes place for a specific choice of the parameters.}
\end{figure}
\begin{figure} \label{k-l-3}
\centering
\begin{tabular}{cc}
{(e)} \includegraphics [scale=0.8, angle=0]{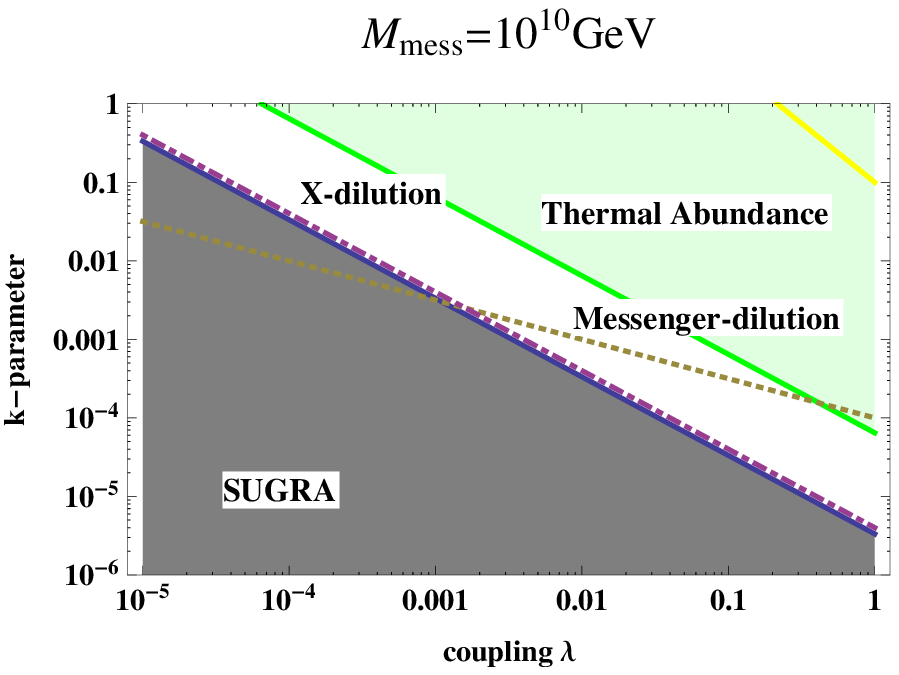}  &
{(f)} \includegraphics [scale=0.8, angle=0] {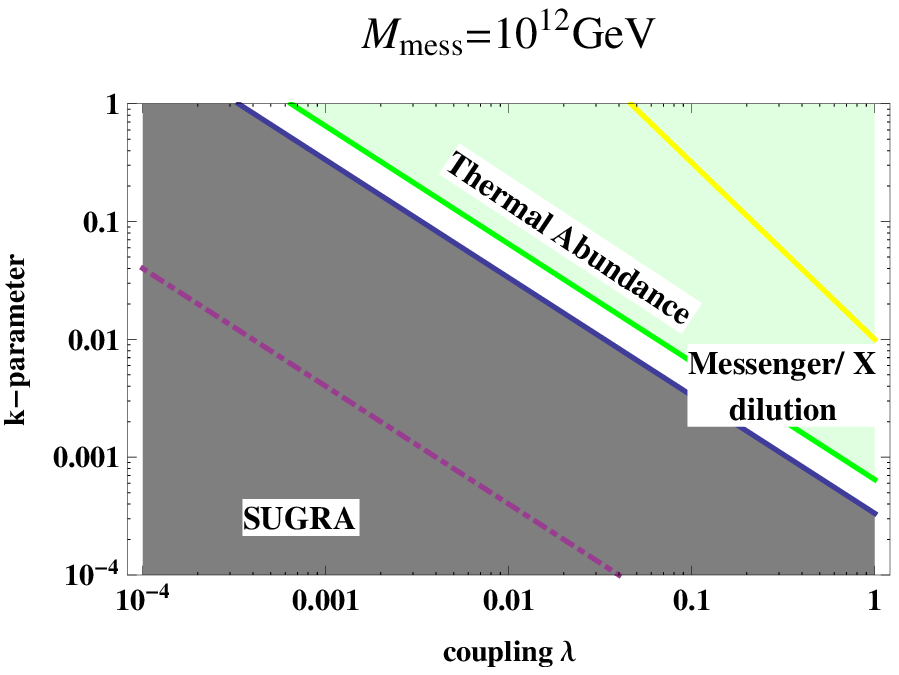}  \\
\end{tabular}
\caption{The same with figure 5 contour plots for large messenger mass $M_\text{mess}$. Also in this contour there is no (yellow) area where $\Omega_{3/2}h^2 \leq 0.11$ is possible without late entropy production.  Here, there is another possible source of dilution: the messenger fields apart from the spurion. The yellow  line in the upper right corner gives $\Omega_{3/2}h^2 = 0.11$. However, dilution by messenger decays requires special couplings and it overlaps the area where generically spurion dilutes. Due to the large messenger masses, we mention here that all (a-f) contours give the helicity $\pm1/2$ gravitino abundance for reheating temperatures $T_\text{rh}> M_\text{mess}$. If $T_\text{rh}< M_\text{mess}$  the gravitino abundance is simply given by the expression (\ref{OMSSM}).}
\end{figure}
The total gravitino abundance is illustrated in the contour plots of figures 5, 6 and 7. All the plots give the $\Omega_{3/2} h^2$ for reheating temperatures higher than the messenger mass. Otherwise there are no thermal excitations of the messenger fields, $R$-symmetry is not restored and the goldstino production is sourced only by the MSSM gauge superfields at $T_\text{rh}$, i.e. the relic abundance is given by the expression (\ref{OMSSM}).

Generally, a thermal messenger population overproduces goldstinos apart from a parameter area (with yellow colour) where it is $\lambda k\ll 1$ and the messenger scale is low, $M_\text{mess}<10^6$ GeV. In the yellow areas of panels (a) and (b), figure 5, the goldstino yield from messenger scatterings is inadequeate to explain the observed relic density of the dark matter. Hence the messenger contribution to $\Omega_{3/2}$ is subdominant and the gravitinos are mainly generated by the MSSM superfields. The reheating temperature can be very high -the only constraint comes from the helicity $\pm3/2$ component- and the gravitinos are generated at the moment that the $R$-symmetry breaks. The gravitino relic density is given by the expression (\ref{ORs}).  The parameters $\lambda, \, k$ and $\Lambda_*$ can be chosen such that $\Omega_{3/2} h^2=0.11$ even when the gluino mass is $m_{\tilde{g}}>1$ TeV. Due to the smallness of the $\lambda\cdot k$ product the gravitino mass, $m_{3/2}\sim 10^{-13}M_\text{mess}/(\lambda k)$, is large: $1\, \text{GeV}< m_{3/2}\lesssim 100$ GeV. We mention that gravitino mass values $m_{3/2}<10$ GeV can be reached for messenger mass scales $M_\text{mess}<10^5$ GeV which are rather constrained by the LHC data.

In the greatest part of the parameter space of the contour plots (a) to (f) the gravitino is overabundant (mostly thermal) due to the messenger scatterings. However, this parameter space is potentially  not fully excluded. The entropy production due to the spurion decay makes a large part of the parameter space viable. There, the dilution by the spurion takes place naturally since it is offset because of the $R$-symmetry restoration and its oscillations are not damped. The spurion oscillations are sizable at the left side of the red dashed line due to smallness of the spurion mass, $m_X(T_{\not{R}})<H(T_{\not{R}})$, and at the right side of the blue dotdashed line due to the disappearance of the non-relativistic messengers from the thermal plasma, $M_\text{mess}(T_{\not{R}})>T_{\not{R}}$. A tuning of the parameters is required in order the gravitino abundance, given by the expression (\ref{OSpu}), to be the correct one. According to \cite{Ibe:2006rc, Hamaguchi:2009hy, Fukushima:2012ra} there are regions in the parameter space for gravitino mass between 10 MeV and 1 GeV where the observed dark matter density can be explained by the decay of the spurion. It is also possible that a dilution caused by the decay of the lightest messenger to take place albeit, as we commented, this is a rather restricted case.

\section{Separate mediation and $R$-breaking sectors}
The models that we have studied until now share a distinct feature: the $R$-symmetry restoration takes place only if the messenger fields have been thermalized. However, it is possible the sector which mediates the supersymmetry breaking not to coincide with the sector responsible for the $R$-symmetry breaking. This is the case if the messengers masses are not controlled by the vev of the spurion field. In particular for $X$ independent messenger masses and $M_\text{mess}\gg T_{\not{R}}$ the goldstino abundance from thermal processes may be given solely by the MSSM sector.
We can sketch an example of a model which implements such a scenario. The superpotential
\begin{equation} \label{WR}
W=FX+\lambda X \phi_1\bar{\phi_2} +\lambda X \phi_2\bar{\phi_1} +M_\text{mess} \left( \phi_1\bar{\phi}_1 +  \phi_2\bar{\phi}_2 \right) + \delta W_h(X,...),
\end{equation}
breaks supersymmetry due to the nonzero vev of $F_X$ component of the spurion field, it mediates the breaking to the observable sector via the $\phi_i$ and $\bar{\phi}_i$  fields which are charged under the SM gauge group and finally breaks the $R$-symmetry via the non-zero vev of the spurion i.e. $\left\langle X \right\rangle=X_0+\theta^2 F$ with $\lambda X_0 \ll M_\text{mess}$. The $\delta W_h$ includes the hidden sector dynamics that stabilize the spurion in the $X_0 \neq 0$ $R$-violating vev. At finite temperature the spurion field can be driven to the origin, $X(T>T_{\not{R}})=0$, due to the $\delta W_h$ dynamics and not due to the messenger fields that stay out of the thermal equilibrium for $T_\text{rh}<M_\text{mess}$. If the goldstino production from the possible hidden sector thermalization is suppressed then the goldstinos are generated at the temperature $T_{\not{R}}$ by the MSSM vector superfields with relic density given by the expression (\ref{OgA}), see figure 3. However, we should mention that the demo model (\ref{WR}) generates a "little" hierarchy between the sfermion and gauginos soft masses due to the fact that not all the messenger fields couple with the spurion field $X$.

\section{Collider and cosmological constraints on the gauge mediation parameters}
\subsection{LHC}
The touchstone of the TeV scale supersymmetry is the ground based experiments. The Large Hadron Collider is nowadays the leading experiment shedding "light" on distances less than $10^{-15}$ cm. Positive or negative experimental evidences for spartiles can support or constrain further the supersymmetric theories. In the framework of specific models these experimental results can, in turn, be translated into constraints on the mass and interactions of gravitino  which is an undetectable particle - thus a dark matter candidate.  

The LHC can indeed discover or give hints for the nature of the dark matter. The dark matter might be produced in the collider and, being stable and (sub)weakly interacting, it will carry away a substantial amount of energy, the so-called missing energy. The gravitino dark matter, being the LSP, can be produced by the NLSP decays. Measurements of the NLSP lifetime and mass can actually reconstruct the gravitino mass, $m_{3/2}$, and the supersymmetry breaking scale $(\sum_i | \left\langle F_i \right\rangle |^2)^{1/2}$. According to eq. (\ref{ssmd}) the NLSP lifetime is  given roughly by 
\begin{align} \label{tnlsp}
c\tau_\text{NLSP} 
\begin{split}
& \approx 48\pi \frac{m^2_{3/2}M^2_\text{Pl}}{\tilde{m}^5_\text{NLSP}} = 16\pi \frac{F^2}{\tilde{m}^5_\text{NLSP}}
\\
& \approx 3 \,\text{m}\, \frac{1}{\lambda^2 k^2} \left(\frac{M_\text{mess}}{10^7\text{GeV}} \right)^2 \left(\frac{100\, \text{GeV}}{\tilde{m}_\text{NLSP}} \right)^5\,.
\end{split}
\end{align}
Collider-stable NLSPs imply a relatively heavy gravitino while prompt decaying NLSPs a light one (appearing as missing energy). 

Despite the entire absence of sparticle signals some assumptions about the messenger mass can be inferred.
Indeed the discovery of a higgs-like boson in the mass region around 125-126 GeV by the CMS and ATLAS teams has important implications for supersymmetry. Notably, this mass is relatively large for the MSSM and large radiative corrections from stop/top loops are needed. Such contributions can arise through stops in the ${\cal O}(10)$ TeV mass range or through lighter stops with maximal mixing \cite{Casas:1994us, Carena:1995bx, Haber:1996fp}. Naturalness promotes the second option (recent discussions on the fine tuning issues can be found in ref. \cite{Evans:2013kxa, Lalak:2013bqa}). A heavy Higgs with light stops can be obtained for large $A$-terms. In the absence of additional interactions $A$-terms are generated through the renormalization group equation of the MSSM driven predominantly by the gluino mass. 
This requires a large messenger scale $M_\text{mess} \gtrsim 10^{10}$ GeV and a heavy gluino $m_{\tilde{g}} \gtrsim 3$ TeV \cite{Draper:2011aa}. Hence, a minimal model of gauge mediation seems to suggest heavy messengers. According to the figure 7 the gravitino for heavy messengers can be the dark matter of the universe if a late entropy production, either by the messengers or the spurion, has taken place. It is interesting to notice that for such large messenger scale the dilution caused by the spurion decay takes place automatically. Also, a heavy gluino does not modify the initial (before the dilution) gravitino abundance which is thermalized - given that $T_\text{rh}>M_\text{mess}$. 

On the other hand, it is possible to generate large $A$-terms at low messenger scales through superpotential interactions between MSSM and messenger superfields \cite{Craig:2012xp, Evans:2013kxa}. Such interactions generate sizable $A$-terms already at the messenger scale while not generating over-large one-loop $m^2_H$ soft masses for the Higgses. This happens by taking into account a one-loop negative contribution to $m^2_{H_u}$ that scales as $\sim\bar{\Lambda}/M_\text{mess}$. For low messenger scale, $M_\text{mess}\sim \bar{\Lambda}$, this contribution is importrant. A direct cosmological implication of this MSSM-messenger mixing is that it renders the messengers rather shortlived ($\Gamma_y \neq 0$) to dominate the energy density of the universe. Moreover, the low messenger scale opens the parameter space towards smaller values of $\lambda k$  where the thermal cross section of goldstino production from messenger thermal scatterings decreases (\ref{sigmas}). For sufficiently small coupling $\lambda$ and parameter $k$ the $Y^\text{mess(sc)}_{3/2}$ can be subdominant and, due to the MSSM-messenger mixing the goldstino yield from the messenger decays $Y^\text{mess(dec)}_{3/2}$ is negligible. Given the cosmological constraint $\Omega_{3/2}h^2 \leq 0.11$, the gravitino will be basically produced from scatterings with the MSSM plasma at the temperature $T_{\not{R}}$ where the $R$-symmetry breaks -given that the helicity $\pm3/2$ gravitino component is underpopulated, see figures 3 and 5. This is an example of gravitino dark matter production due to a symmetry violation. In addition, once messenger scale is indicated then the $\lambda$ and $k$ parameters could be also probed.

Other models exist that may lead to different conclusions. It is not the purpose of this work to present an exhaustive study all the GMSB models that aim to fit the LHC data. Here we briefly alluded to some general features of the GMSB phenomenology in the LHC era and exposed how the collider data can be directly connected, reconciled or contrasted with the gravitino cosmology.

Finally, we note that, according to the findings of this work, cosmology implies a different kind of relation between the GMSB messenger scale and the gravitino mass. For a universe reheated only once (e.g. from the inflaton field) to high temperatures then the messenger scale is low, $M_\text{mess}<10^6$ GeV and the gravitino mass in the range  $1\, \text{GeV} < m_{3/2} \lesssim 100$ GeV. Otherwise, the gravitino is overabundnat, see figure 5. A heavy gravitino makes the NLSP a collider stable particle -it actually decays outside the solar system. In the case that the spurion field produces late entropy then the gravitino can be lighter. Of course all these constraints apply for $T_\text{rh}>M_\text{mess}$. The reheating temperature of the universe is suggested to be that high, $T_\text{rh}>M_\text{mess}$, by two reasons: the thermal selection of the supersymmetry breaking minimum \cite{Dalianis:2010pq} and the theory of thermal leptogenesis. As we will discuss in the next subsection the gravitino mass and the NLSP lifetime have additonal, important implications for cosmology.

\subsection{Further cosmological constraints}
\itshape{Large scale structure} \normalfont
\\
Cosmological observations strongly indicate that the dark matter of the universe is cold \cite{Ade:2013lta}. The gravitino is a light particle and if too light it can behave like warm dark matter. The gravitinos in the early universe are relativistic and their momentum redshifts like $p\propto a^{-1}$. However, non-thermally produced gravitinos by the decay of the spurion may be still relativistic at late times, for their production  takes place at low temperatures $T\sim$ MeV - GeV. The free-streeming lenght of gravitinos has to be consistent with the observational bound $\lambda_{fs} \lesssim {\cal O}(100)$ kpc from the Lyman $\alpha$ forest data \cite{Boyarsky:2008xj}. 
\begin{equation} \label{Ly}
\lambda_{fs} \sim 100 \,\text{kpc}\,\left(\frac{15}{g_*} \right)^{1/4}\left(\frac{100\,\text{MeV}}{m_{3/2}} \right) \left(\frac{m_X}{500\,\text{GeV}} \right) \left(\frac{16\,\text{MeV}}{T_\text{spur}} \right).
\end{equation}
A supplementary bound comes from the reionization epoch \cite{Ade:2013lta}. These bounds constrain the gravitino mass from being too small and can be applied on our results. According to the figures 5, 6 and 7 there is a parameter region (upper right corner, i.e. where $\lambda k \rightarrow$  1) where the gravitino is produced non-thermally from the spurion decay. The low messenger scale, $M_\text{mess} \lesssim 10^8$ GeV, and the relatively large values of the parameters $\lambda$ and $k$ correspond to a light gravitino. Hence, this part of the parameter space can be excluded as a viable choice due to the observational bound (\ref{Ly}).
\\
\itshape{BBN} \normalfont
\\
On the other hand, a heavy gravitino although compatible with the large scale structure constraints it makes the $\tau_\text{NLSP}$ large  (\ref{tnlsp}) rendering the NLSP decay potentially dangerous for ordinary nucleosynthesis predictions. The BBN constrains the $\tau_\text{NLSP}$ to be up to $\sim10^8$ sec if the final-state particles are photons and this upper bound decreases significantly, nearly 10 orders of magnitude, for hadronic NLSP decays. A heavy gravitino $1\, \text{GeV} \lesssim m_{3/2} \lesssim 100$ GeV is predicted when there is no entropy production other than the very initial one; differently, the gravitino relic density exceeds the observational bound $\Omega h^2_{3/2} \lesssim 0.11$ (see figure 5). When the gravitino mass exceeds a few GeV then cosmology requires either a comparatively heavy NLSP or/and an NLSP with small hadronic branching ratio or/and with small relic abundnace. In gauge mediation the NLSP is generically a neutralino or a stau and in some particular cases a sneutrino. The cosmology of a heavy stable gravitino, $1\, \text{GeV} < m_{3/2} \lesssim 100$ GeV, can be viable and indicates towards specific gauge mediation supersymmetry breaking schemes. A recent discussion can be found in the ref. \cite{Roszkowski:2012nq}.
\\
\itshape{Leptogenesis} \normalfont
\\
The fact that the goldstino abundance is nearly independent of the reheating temperature with its yield being dominated either by the temperature $T_{\not{R}}$ or $M_\text{mess}$ allows the thermal leptogenesis scenario to be realized without violating the bound $\Omega_{3/2} h^2 \lesssim 0.11$. In the leptogenesis scenarios the baryon asymmetry, $\eta_B \simeq 6\times 10^{-10}$, is generated in the out-of-equilibrium decay of right handed neutrinos and sneutrinos in the very early universe \cite{Davidson:2002qv, Davidson:2008bu}. The right handed neutrinos can be generated by scatterings in the thermal bath for
\begin{equation}
T_\text{rh}\gtrsim M_1 \sim 10^9 \,\text{GeV}\,,
\end{equation}
where $M_1$ the mass of the lightest of the heavy neutrinos. In this work we demonstrated that such high temperatures are allowed in the GMSB scenarios even in the absence of late entropy production given that the messenger scale is low enough, namely, $M_\text{mess}<10^6$ GeV. If, on the other hand, the messenger scale is not that low then leptogenesis can still take place while satisfying the dark matter density bound due to dilution caused by the decay of the spurion field. The insensitivity of the helicity $\pm1/2$ gravitino relic density to the reheating temperatures for $T_\text{rh}>M_\text{mess}$ can help in achieving the correct dilution magnitude for the gravitino abundance without the washing-out of the baryon asymmetry.

\section{Conclusions}
The gravitino is a hypothetical particle predicted by a well motivated theory, the supersymmetry, which currently is being tested at the LHC. In the gauge mediation supersymmetry breaking (GMSB) theories the gravitino is the lightest supersymmetric particle and if $R$-parity is conserved then it can be part of the dark matter of the universe. The gravitino relic density depends on its couplings and the reheating temperature of the universe. 
For reheating temperatures below the messenger scale the gravitino is produced solely by the MSSM particles. For reheating temperatures above the messenger scale the messengers have a dominant contribution and we actually find that the gravitino can easily attain a thermal equilibrium distribution. High reheating temperatures generically restore the $U(1)_R$ symmetry of GMSB sectors suppressing the gravitino production from the MSSM superfields. Furthermore, the GMSB spurion $X$ acquires a thermal vev about the origin of the field space and if its oscillations amplitude is large enough it can produce entropy at late times. 

The gravitino can have a relic abundance $\Omega_{3/2} \leq \Omega_\text{DM}$ for different values of the parameter space which controls the gravitino yield. If the messenger scale is low, $10^5 \text{GeV} \lesssim M_\text{mess} \leq 10^6$ GeV, and the messenger fields have superpotential couplings with the MSSM superfields then the gravitino yield from the thermalized messenger scatterings and decays may not be able to saturate the $\Omega_\text{DM}$ bound. This is the case when GMSB messenger superpotential coupling $\lambda$ is sufficiently small. This is actually suggested by the conditions of the thermal selection of the supersymmetry breaking vacuum \cite {Dalianis:2010pq}. Moreover, the gravitino has to be heavy, $10\, \text{GeV} \lesssim m_{3/2} \leq {\cal O}(100)$ GeV hence, we conclude that there is an extra hidden sector, $Z$, that dominantly contributes to the supersymmetry breaking but has a subdominant, gravitational contribution to the soft masses.
It is natural to expect that this hidden sector is characterized by an exact U$(1)_R$ symmetry \cite{Nelson:1993nf}. The $R$-symmetry, on the one hand, makes the $Z$-vacuum an enhanced symmetry point and, on the other, implies that the $Z$-sector contributes mainly to the sfermion soft masses.

The inadequate gravitino production from the thermalized messengers with $M_\text{mess}\leq 10^6$ GeV implies that the gravitinos are  produced mainly by the MSSM sector. The MSSM superfields generate gravitinos dominantly at the temperature, $T_{\not{R}}$, which is the characteristic temperature of the transition to the $R$-violating phase \cite{Dalianis:2011ic}. The gravitino relic density can have the observed dark matter density, $\Omega_{3/2}h^2 \simeq 0.11$, in a quite constrained but not negligible part of the parameter space.  For that low messenger scales the GMSB spurion can reach the zero temperature minimum without producing late entropy.

When the messenger scale is large the gravitino yield from the thermalized messengers increases.  The gravitino is thermally over-produced. However, a large messenger scale implies a large vev for the GMSB spurion, $X_0$, and it is found that generically the spurion field dilutes the thermal plasma and produces non-thermal gravitinos. It is possible the dilution and the non-thermal gravitino yield to have an acceptable size \cite{Fukushima:2012ra}.

It is interesting that the current GMSB phenomenological models discuss the cases of low and large messenger scale. The input from particle phenomenology is essential for describing the cosmology of the gravitino in a more concrete and consistent way. In this work we have shown that gravitino cosmology can be compatible with high reheating temperatures with the upper bound imposed by the yield of the helicity $\pm3/2$ gravitino component. Therefore, the thermal selection of the supersymmetry breaking vacuum \cite {Dalianis:2010pq} and the leptogenesis scenario  can be realized without facing the gravitino overproduction problem. The following years of the LHC data will hopefully shed "light" on the physics of dark matter.

\section*{Acknowledgments}
\vspace*{.5cm}
\noindent 
It is a pleasure to thank Alex Kehagias for discussion and comments.

\appendix
 
\section{Thermal restoration of the $U(1)_R$ symmetry}

\subsection{Exact $R$-symmetry}
The connection between a global $U(1)_R$ symmetry and the supersymmetry breaking is well established \cite{Nelson:1993nf}: $R$-symmetry  for generic superpotentials is a necessary condition for supersymmetry breaking in the true vacuum and a spontaneously broken $R$-symmetry a sufficient one. We expect $R$-symmetry to be violated by different sources that may restore supersymmetry but not in a nearby vacuum.  These sources may be $1/M_\text{Pl}$ suppressed dimension-five operators in the superpotential or a constant term that cancels the cosmological constant in the vacuum.

In the global supersymmetric limit the $U(1)_R$ symmetry can be an exact symmetry. A model that implements this behaviour has the minimal superpotntial (\ref{min-3}) and K\"alher the (\ref{KSBR-3}) with $\epsilon_4=1$ and $\epsilon_6<0$. This K\"ahler potential can originate from an $R$-symmetric O'Raifeartaigh-like sector that breaks spontaneously the $R$-symmetry, see for example ref. \cite{Shih:2007av}.
The $R$-symmetry is a symmetry of the vacuum when $X=0$ and breaks spontaneously when $X\neq 0$. It is restored due to thermal effects at the temperature \cite{Dalianis:2010yk, Dalianis:2011ic} 
\begin{equation} \label{TR-9}
T_{\not{R}}= \frac{4}{\sqrt{N}} \frac{F}{\lambda \Lambda_*}.
\end{equation}
For the minimal case of a $\bold{5}+\bold{\bar{5}}$ messenger sector, i.e. the $\phi$, $\bar{\phi}$ messenger quarks and leptons form a single complete $SU(5)$ representation, it is $N=5$. We can have additional $SU(5)$ multiplets that couple to the spurion $X$ field preserving the gauge unification. 
For large messenger scales $N$-values as large as $\sim50$ are allowed \cite{Giudice:1998bp}.
The $T_{\not{R}}$ can be quickly estimated: the spurion has negative squared mass $4F^2/\Lambda^2_*$ at the origin and receives thermal corrections $N\lambda^2 T^2/4$ from the messenger fields.

\subsection{Approximate $R$-symmetry}
Two basic examples of gauge mediation are the following \cite{Kitano:2006wz, Murayama:2006yf}:
\begin{equation}
W=FX+\lambda X\phi\bar{\phi} + c, \quad \quad W=FX+\lambda X\phi\bar{\phi} - M\phi\bar{\phi} 
\end{equation}
with K\"ahler, $K=|X|^2-|X|^4/\Lambda^2_*$ and  $N$ number of messenger fields $\phi$ and $\bar{\phi}$ in the fundamental representation. Although these theories break the $U(1)_R$ explicitly they are approximately $R$-symmetric at high temperatures. Indeed, for high enough temperatures the $R$-violating terms $c$ and $M\phi\bar{\phi}$ are negligible and an approximate $R$-symmetry restoration takes place. This can be seen from the evolution of the thermal average value for the $R$-charged $X$ field \cite{Dalianis:2010yk}: 
\begin{equation} \label{X-min09}
X^{(c)}_{min}(T)=\frac{4\frac{c}{M^2_\text{Pl}} F -\frac{2Fc}{3\Lambda^2M^2_\text{Pl}}T^2}{8\frac{F^2}{\Lambda^2_*}+\frac{N}{2}\lambda^2T^2}, \quad \quad X^{(M)}_{min}(T)=\frac{\frac12 M\lambda T^2}{8 \frac{F^2}{\Lambda^2_*}+\frac{N}{2}\lambda^2 T^2}
\end{equation}

The $R$-symmetry breaking scale is the vev $\left \langle X \right\rangle \equiv X_0$. For the first case of gravitational stabilization the vev is the $X^{(c)}_0 =c\Lambda^2/(2FM^2_\text{Pl})$. 
For the second model, after the translation $X\rightarrow M/\lambda -X$  it is $X^{(M)}_0 = M/\lambda$ thus, the $X_0$ represents the scale of $R$-breaking.

According to (\ref{X-min09}) the thermal average value tends to restore the $R$-symmetry. We can parametrize the degree of the $R$-symmetry breaking by defining the parameter $b_R$:
\begin{equation}
b_R(T)\equiv \frac{X(T)}{X_0}\,.
\end{equation}
Temperatures higher than the cut-off scale are not expected (for $\Lambda_* \gtrsim 10^{-4}M_\text{Pl} $) since a thermal equilibrium cannot be achieved. Thus, for the case of gravitational stabilization the second term at the numerator (\ref{X-min09}) is negligible. The parameter $b_R$ is given, for both cases, from the expression
\begin{equation} \label{b1-9}
b_R(T)=\frac{\left( 4 \frac{F}{\Lambda_*} \right)^2 }{ \left( 4 \frac{F}{\Lambda_*} \right)^2 + N \lambda^2 T^2}
\end{equation}
Utilizing the definition (\ref{TR-9}) of the $T_{\not{R}}$ we can recast the (\ref{b1-9}) into the simpler form
\begin{equation} \label{T0-9}
b_R(T) = \frac{1}{1+\left(\frac{T}{T_{\not{R}}} \right)^2}\,.
\end{equation}
Obviously, when $T\rightarrow 0$ the $R$-symmetry breaking scale takes its maximum value, i.e. the zero temperature one, and when $T \rightarrow \infty $ the $R$-symmetry is restored. In other words, the $b_R(T)$ parametrizes the $R$-symmetry breaking scale at finite temperature with respect to the zero temperature scale. 
For the case of spontaneous breakdown of the $R$-symmetry, discussed in the previously, the parameter $b_R(T)$ takes, approximately, the discrete values: 
\begin{equation}
b_R(T>T_{\not{R}})=0
\quad \quad \text{and} \quad \quad
b_R(T<T_{\not{R}})=1\,.
\end{equation}
From the expression (\ref{T0-9}) we see that $b_R(T_{\not{R}})=0.5$. 
Also, from (\ref{T0-9}) we can re-derive the temperatures that the supersymmetry breaking vacua form, $T_X$, firstly given in ref. \cite{Dalianis:2010yk,Dalianis:2010pq}. The $T_X$ controls the thermal selection of the supersymmetry breaking vacuum. It corresponds to the temperature that the minimum at the $X$-direction crosses the tachyonic boundary $X=\sqrt{F/\lambda}$. Hence, 
\begin{equation}
X(T_X)= \sqrt{F/\lambda}=b_R(T_X)X_0
\end{equation}
which gives the following values for the parameter $b_R$:
\begin{equation}
b_R( T^{(c)}_X )=2\frac{F\, M^2_\text{Pl}}{c\Lambda^2_*}\sqrt{\frac{F}{\lambda}}\, , \quad \quad \quad b_R(T^{(M)}_X)=\frac{\sqrt{\lambda F}}{M}.
\end{equation}
It has to be $b_R( T^{(c)}_X ), \, b_R( T^{(M)}_X )\ll 1$ and from (\ref{T0-9}) we take
\begin{equation}
T^2_X \simeq \frac{8}{N}\frac{c}{\lambda M^2_\text{Pl}} \sqrt{\frac{F}{\lambda}}\, \quad \quad \text{and} \quad\quad T^2_X \simeq \frac{16}{N} \frac{F M}{\lambda^2 \Lambda^2_*}\sqrt{\frac{F}{\lambda}}\, 
\end{equation}
which are the temperatures derived in ref. \cite{Dalianis:2010yk,Dalianis:2010pq}. We also note that  $T_X>T_{\not{R}}$.

\section{Gravitino production from the messenger sector}
The gravitino yield $Y_{3/2}$, i.e. the number to entropy ratio, is found by solving the Boltzmann equation describing the gravitino production
\begin{equation} \label{bolz}
\frac{dY_{3/2}}{dT}=\frac{1}{sHT}\left(\sum_i\left\langle \Gamma_{i\rightarrow \psi_G+...}\, n_i \right\rangle + \sum_{i,j} \left\langle \sigma_{i+j\rightarrow \psi_G+...} \, v_{ij} \, n_i\,  n_j \right\rangle \right)\,.
\end{equation}
The scattering processes of messenger fields contributing to goldstino production are $\chi_{\bar{\phi}}+V^a_\mu \rightarrow \phi + \psi_G$,  $\phi+ \chi_{\bar{\phi}} \rightarrow V^a_\mu + \psi_G$, $\chi_{\bar{\phi}}+\lambda^a \rightarrow \chi_{\bar{\phi}} + \psi_G$, $\phi + \lambda^a \rightarrow \phi^* + \psi_G$, $\chi_{\phi} + \chi_{\bar{\phi}} \rightarrow \lambda^a + \psi_G$ and $\phi + \phi^* \rightarrow \lambda^a + \psi_G$.
The corresponding cross sections to goldstinos given in ref. \cite{Choi:1999xm} read
\begin{equation} \label{cs1}
\sum_{A, \, B,\, B'}  \sigma(A_\text{mess}+ B_\text{MSSM}   \rightarrow A'_\text{mess} + \psi_G)   =\xi \lambda^2 k^2\frac{2(2s^2-3sM^2_\text{mess}+M^4_\text{mess})+s(s-2M^2_\text{mess})\log(\frac{s^2}{M^4_\text{mess}}) }{s(s-M^2_\text{mess})^2}
\end{equation}
and 
\begin{equation} \label{cs2}
\frac{1}{2} \sum_{A,\, A', \, B} \sigma(A_\text{mess}+A'_\text{mess} \rightarrow B_\text{MSSM} + \psi_G) \simeq\, \xi \lambda^2 k^2\frac{2[s(s-4M^2_\text{mess})]^{1/2}+M^2_\text{mess}\log\left(\frac{s-2M^2_\text{mess}-[s(s-4M^2_\text{mess})]^{1/2}}{s-2M^2_\text{mess}+[s(s-4M^2_\text{mess})]^{1/2}} \right)}{s(s-4M^2_\text{mess})}\, ,
\end{equation}
where $A_\text{mess} = \phi, \, \chi_\phi$ and $B_\text{MSSM}= \lambda^a,\, V^a_\mu$ are respectively messenger and MSSM gauge superfield components.

The decay width of a spin 1/2 messenger to a scalar messenger plus a goldstino due to the Lagrangian interaction $\delta {\cal L} = \lambda k \, \psi_G \, \chi_{\bar{\phi}}\, \phi$ reads \cite{Drees:2004jm}
\begin{equation} 
\Gamma_\lambda (\chi_{\bar{\phi}} \rightarrow  \phi \, \psi_G)= 4\times\frac{k^2 \lambda^2} {16\pi} M_{\chi_{\bar{\phi}}} \left(1- \frac{M^2_\phi}{M^2_{\chi_{\bar{\phi}}}} \right)^2.
\end{equation}
The fermion messenger mass is $M_{\chi_{\bar{\phi}}}=\lambda X_0 \equiv M_\text{mess}$ and the squared mass of the scalar messengers is $M^2_\phi= M^2_\text{mess}  \pm \lambda F$. Hence, the decay rate is recast into
\begin{equation}
\Gamma_\lambda (\chi_{\bar{\phi}} \rightarrow  \phi \, \psi_G) = \frac{1}{4\pi} \lambda^4 k^4 \frac{F^2}{M^3_\text{mess}}.
\end{equation}
\begin{figure} \label{}
\centering
\begin{tabular}{cc}
{(a)} \includegraphics [scale=0.8, angle=0]{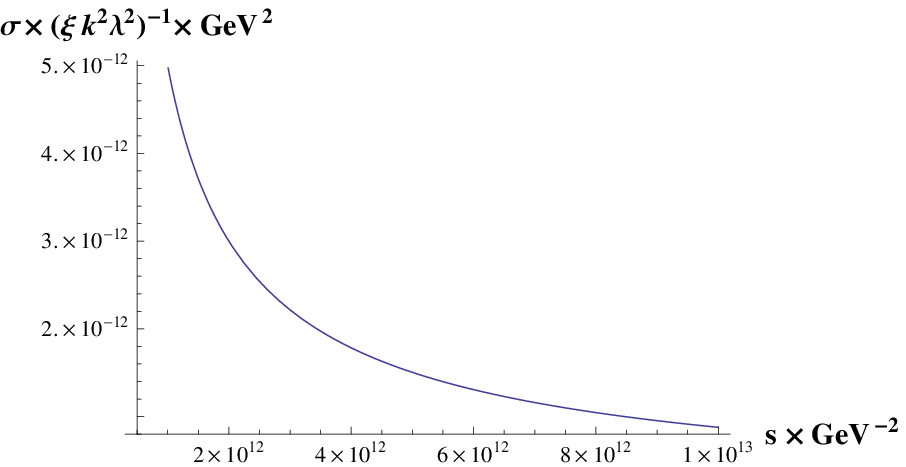}  &
{(b)} \includegraphics [scale=0.8, angle=0] {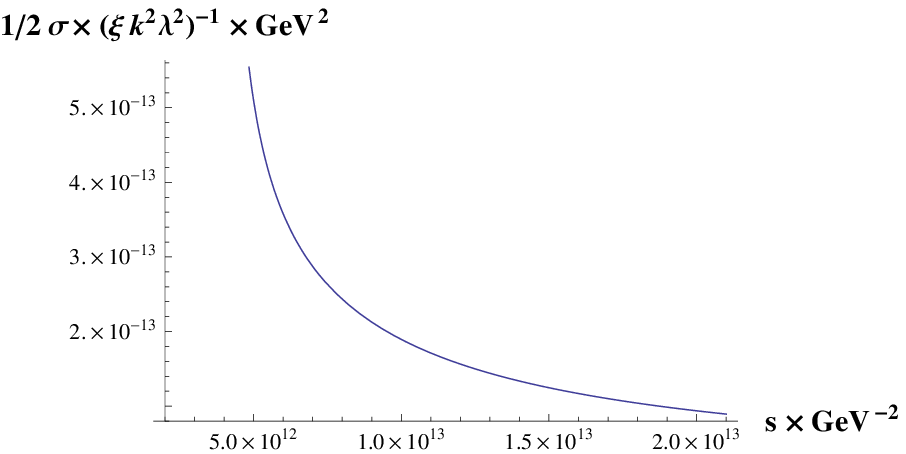}  \\
\end{tabular}
\caption{The scattering cross sections (\ref{cs1}) and (\ref{cs2}) of thermalized messengers to goldstinos as a function of the center of mass energy squared. The figures show the fast growth of the cross section close to the messenger mass squared and the $1/s$ fall at high energies. Here the exemplified messenger mass is $M_\text{mess}=10^6$ GeV. }
\end{figure}
\\
\itshape Messenger Scatterings \normalfont
\\
The thermally averaged product appearing in (\ref{bolz}) is found in \cite{Kolb:1990vq} and takes the form
\begin{equation}
\left\langle \sigma_{i+j\rightarrow \psi_G+...} \, v_{ij} \, n_i\,  n_j \right\rangle = \frac{T}{8\pi^4}\int^\infty_{(M_i+M_j)^2} ds \, \sqrt{s} \,K_1 \left(\frac{\sqrt{s}}{T}\right) \, |\vec{p}|^2 \sigma_\text{tot}(s)
\end{equation}
with $\sigma_\text{tot}$ the sum of the scattering cross sections (\ref{cs1}) and (\ref{cs2}), $K_1$ the modified Bessel function of the second kind of degree one, and $|\vec{p}|$ the absolute value of the incident $i$ particle in the center of mass frame. The scattering yield is given by the expression
\begin{align}
Y^\text{mess(sc)}_{3/2}
\begin{split}
& = \int^{T_\text{rh}}_{T} \frac{\left\langle \sigma_{i+j\rightarrow \psi_G+...} \, v_{ij} \, n_i\,  n_j \right\rangle}{s(T)H(T)T} dT \\
& = \frac{T^6}{16 \pi^4\,s(T)H(T)T} \int^\infty_{x_\text{rh}}dx \, x^3 K_1(x) \int^{xT_\text{rh}}_{M_i+M_j} d\sqrt{s}\,  \frac{4|\vec{p}|^2}{s}\,\sigma_\text{tot}(s) \,
\end{split}
\end{align}
with $x=(M_i+M_j)/T$. When one of the incident particles is a messenger then $M_i=M_\text{mess}$ and $M_2=0$ we have the yield
\begin{equation} \label{Yscat}
\left. Y^\text{mess(sc)}_{3/2} \simeq \frac{124 \pi}{15} \frac{T^6}{16\pi^4\,s(T)H(T)T} \right|_{T=M_\text{mess}} \xi \, \lambda^2 k^2 \frac{1}{M_\text{mess}}\, =\, 1.3 \times 10^{-5} \lambda^2 k^2 \left(\frac{270}{g_*} \right)^{3/2}\frac{M_\text{Pl}}{M_\text{mess}}
\end{equation}
\\
Gravitinos reach equilibrium abundnace when $\Gamma_{G\phi}>H$, see eq. (\ref{goldeq}), for at least one expansion time. The relic abundance of thermal gravitinos is given by the expression \cite{Kolb:1990vq}
\begin{equation}
Y^\text{eq}_{3/2,\, \infty} \simeq \frac{0.278\times 3/2}{g_*(T\sim M_\text{mess})} \left(1-\text{exp}\left[-\int_0^\infty \frac{\Gamma_{G\phi}}{HT'}dT' \right]  \right)
\end{equation}
\\
\itshape Messenger Decays \normalfont
\\
The gravitino yield from the decays of the thermalized messengers is given by
\begin{equation} \label{ydec}
Y^{\text{mess(dec)}}_{3/2} (T) 
 =\int^{T_\text{rh}}_{T}dT\frac{\sum_l n^\text{eq}_l \Gamma^\text{mess}_\text{dec} (M_\text{mess}/ \left\langle E \right\rangle) }{s(T)H(T)T} = \int^{T_\text{rh}}_{T}dT\frac{\sum_l (T^3/\pi^2) x^3 K_1(x) \Gamma^\text{mess}_\text{dec} }{s(T)H(T)T}\, ,
\end{equation}
where the sum is over all the messenger particles. The yield is dominated by temperatures $T\sim M_\text{mess}$. The decay width of messengers to goldstinos is given by the expression 
\begin{equation}
\Gamma_\lambda(\chi_\phi \rightarrow \phi\,\psi_G)=4\times \lambda^2 k^2\frac{M_\text{mess}}{16\pi}\left(1-\frac{M^2_\phi}{M^2_{\chi_\phi}} \right)^2=\lambda^2 k^2 \frac{M_\text{mess}}{4\pi}\left(\frac{\lambda F_X}{M^2_\text{mess}}\right)^2=\frac{1}{4\pi} \lambda^4 k^4 \frac{F^2}{M^3_\text{mess}}\,.
\end{equation}
For $\Gamma_\text{dec}^\text{mess}=\Gamma_\lambda$ the yield (\ref{ydec}) takes the value
\begin{align} \label{Ydecay}
Y^{\text{mess(dec)}}_{3/2} 
\begin{split}
 & \left.\left. =\frac{3\pi}{2} \frac{T^4\,\sum_l\Gamma_\lambda}{\pi^2 s(T)H(T)T} \right|_{T=M_\text{mess}}=\frac{135}{4\pi^3}  \frac{\sum_l \Gamma_\lambda}{g_*(T)H(T)}\right|_{T=M_\text{mess}} \\
 & =1.4 \times 10^{-5} \left(\frac{N_{\phi\bar{\phi}}}{10}\right) \left( \frac{270}{g_*}\right)^{3/2} \left(\frac{\lambda k}{10^{-6}} \right)^2\left(\frac{10^6\,\text{GeV}}{M_\text{mess}} \right)^3 \left(\frac{\bar{\Lambda}}{10^5 \,\text{GeV}}\right)^2
\end{split}
\end{align}
where $N_{\phi\bar{\phi}}$ the number of the messengers that decay to lighter messenger superpartners plus the goldstino. It corresponds to the relic abundance
\begin{equation}
\Omega^\text{mess(dec)}_{3/2} \simeq 0.5 \,N_{\phi\bar{\phi}} \left(\frac{10^6\,\text{GeV}}{M_\text{mess}}\right)\left(\frac{\text{GeV}}{m_{3/2}}\right) \left(\frac{m_{\tilde{g}}}{\text{TeV}}\right)^4.
\end{equation}

However, this goldstino yield from messengers decays changes when there are more decay channels. Actually, the goldstino yield from messenger decays cannot exceed the thermal equilibrium value of relativistic messenger fields, $Y^\text{eq}N_{\phi\bar{\phi}} \sim 10^{-3}N_{\phi\bar{\phi}}$, because the goldstino is one of the two daughter particles from the decay of the messenger, the mother particle. Moreover, given the existence of other decay channels of messengers to MSSM particles with decay width $\Gamma_y$ and $\Gamma_m$ the goldstino yield from messenger sector decays is bound to be less than the thermal equilibrium value regardless the value of the product $\lambda\cdot k$ and the messenger mass $M_\text{mess}$. Taking into account the other decay channels of the messenger fields  the goldstino yield from the messenger decays is 
\begin{equation}
Y^\text{mess(dec)}_{3/2} \sim B_{3/2} N_{\phi\bar{\phi}} Y^\text{eq}_\text{mess} 
\end{equation}
where $B_{3/2}=\Gamma_\lambda/(\Gamma_y+\Gamma_m+\Gamma_\lambda +...)$. 
Furthermore it is $K(T=M_\text{mess})\equiv \Gamma_\text{tot}/H \gg 1$. 

The expression (\ref{Ydecay}) has a $M^{-3}_\text{mess}$ dependence and appears to dominate over the goldstino yield from scatterings, (\ref{Yscat}), for $M_\text{mess}\lesssim 10^{6}\text{GeV}$.  However, when there are other messenger decay channels this conclusion is altered. In particular for $\lambda k<10^{-6}$ the product $\lambda k$, according to the ratios (\ref{lambday}) and (\ref{lambdam}) depicted in the figure (\ref{dw}), render the decay rate $\Gamma_\lambda$ negligible compared to the $\Gamma_y$ and $\Gamma_m$. 
 Hence, we find that $B_{3/2}\simeq \Gamma_\lambda/(\Gamma_y+\Gamma_m)<\{10^{-15}-10^{-2} \}$  and it is 
\begin{equation}
Y^\text{mess(dec)}_{3/2}<Y^\text{mess(sc)}_{3/2}\leq Y^\text{eq}_{3/2}\,,
\end{equation} 
for {\itshape any} messenger scale.

\end{document}